\documentclass[a4paper,fleqn,usenatbib]{mnras}

\usepackage{mathptmx}

\usepackage[T1]{fontenc}
\usepackage{ae,aecompl}

\usepackage{graphicx}	
\usepackage{amsmath}	
\usepackage{amssymb}	

\usepackage{booktabs}
\usepackage{physics}
\usepackage{hyperref}

\usepackage{xr}
\title[Surface Dynamics, Equilibria and Lobes of Arrokoth]{Surface Dynamics, Equilibrium Points and Individual Lobes of the Kuiper Belt Object (486958) Arrokoth}

\author[A. Amarante et al.]{
A. Amarante$^{1,2,3}$\thanks{E-mail: andre.amarante@unesp.br}\href{https://orcid.org/0000-0002-9448-141X}{\includegraphics[scale=0.5]{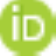}}
and O. C. Winter$^{2}$\thanks{E-mail: othon.winter@unesp.br}\href{https://orcid.org/0000-0002-4901-3289}{\includegraphics[scale=0.5]{orcid_16x16.pdf}}
\\
$^{1}$State University of Mato Grosso do Sul - UEMS, Cassil\^andia, CEP 79540-000, Mato Grosso do Sul, Brazil\\
$^{2}$Grupo de Din\^amica Orbital e Planetologia (GDOP), S\~ao Paulo State University - UNESP,\\ Guaratinguet\'a, CEP 12516-410, S\~ao Paulo, Brazil\\
$^{3}$Simula\c c\~ao Num\'erica Computacional (SONICO), Federal Institute of Education, Science and Technology of S\~ao Paulo - IFSP,\\ Cubat\~ao, CEP 11533-160, S\~ao Paulo, Brazil
}

\date{Accepted 2020 June 11. Received 2020 June 4; in original form 2020 March 8.}

\pubyear{2020}

\hypersetup{draft}
\begin{document}
\label{firstpage}
\pagerange{\pageref{firstpage}--\pageref{lastpage}}
\maketitle

\begin{abstract}
The New Horizons space probe led the first close flyby of one of the most primordial and distant objects left over from the formation of the solar system, the contact binary Kuiper Belt object (486958) Arrokoth, which is composed of two progenitors, the lobes nicknamed Ultima and Thule. In the current work, we investigated Arrokoth's surface in detail to identify the location of equilibrium points and also explore each lobe's individual dynamic features. We assume Arrokoth's irregular shape as a homogeneous polyhedra contact binary. We numerically explore its dynamic characteristics by computing its irregular binary geopotential to study its quantities, such as geometric height, oblateness, ellipticity, and zero-power curves. The stability of Arrokoth Hill was also explored through zero-velocity curves. Arrokoth's external equilibrium points have no radial symmetry due to its highly irregular shape. We identified even equilibrium points concerning its shape and spin rate: i.e., four unstable external equilibrium points and three inner equilibrium points, where two points are linearly stable, with an unstable central point that has a slight offset from its centroid. Moreover, the large and small lobes each have five equilibrium points with different topological structures from those found in Arrokoth. Our results also indicate that the equatorial region of Arrokoth's lobes is an unstable area due to the high rotation period, while its polar locations are stable resting sites for surface particles. Finally, the zero-power curves indicate the locations around Arrokoth where massless particles experience enhancing and receding orbital energy.
\end{abstract}




\section{Introduction}
\label{sec:intro}
The small contact binary (486958) Arrokoth (provisionally designated 2014 MU$_{69}$ and unofficially named `Ultima Thule' by the New Horizons team) is the farthest and most ancient body in the solar system visited by the New Horizons spacecraft. Arrokoth is a member of the cold classical Kuiper Belt objects with low inclinations and near-circular orbits that are remnant materials of the building blocks of the solar system \citep{Delsanti2006}. Thus, our analysis of the preserved data from Arrokoth obtained by the New Horizons planetary probe will be important for understanding the role that primitive Kuiper Belt objects may have played in planetary formation processes. In addition, the small planetesimal Arrokoth is also the first primordial contact binary ever explored in situ.

In the present work, we built two main numerical tools and use the low facet polyhedral model of Arrokoth presented in \citet{Stern2019b} to investigate this system further. To explore the dynamic geophysical environment on each Arrokoth lobe surface specifically, we first use a modified version of \citet{Tsoulis2012}'s code called the \textit{Minor-Gravity package}\footnote{\url{https://github.com/a-amarante/minor-gravity}.} to numerically compute the binary's gravitational potential, as well as its first- and second-order derivatives. Moreover, we also concentrate on the equilibria of Arrokoth. We adopt a \textit{Minor-Equilibria package}\footnote{\url{https://github.com/a-amarante/minor-equilibria}.} to find the location of the equilibrium points and study their topological structures. 

This paper is structured into the following sections. In the next section, we reproduce a three-dimensional (3-D) polyhedral model of Arrokoth in terms of its geometric height. In addition, we discuss its topographic and physical features using their surface tilts. We use our adopted mathematical model to explore the binary gravitational force potential of the Arrokoth contact binary and the results for equipotential curves and lines of force around Arrokoth are presented and discussed in Section \ref{sec:grav}. We also compute the oblateness and ellipticity of Arrokoth's large and small lobes. Section \ref{sec:forc} presents the results of Arrokoth's geometric and physical characteristics presented in Section \ref{sec:prop} in relation to the dynamics of surface particles. In this section, we computed the binary's geopotential surface, surface accelerations, slopes, and escape speed to study the surface stability of each Arrokoth lobe. In Section \ref{sec:eq}, we show the equilibrium points' location, their stability, and the topological structures of Arrokoth and its individual lobes. Additionally, we analyzed the effect of different densities on the dynamic properties of their equilibria. The results associated with the stability of Arrokoth Hill and the return speed, orbital energy and gravity power of the Arrokoth contact binary are also discussed and presented at the end of this section. Finally, in the last section, we provide our conclusions and some perspectives.

\section{Geometric and physical features of the Arrokoth shape model}
\label{sec:prop}
\begin{figure*}
  \centering
  \fbox{\includegraphics[width=18cm]{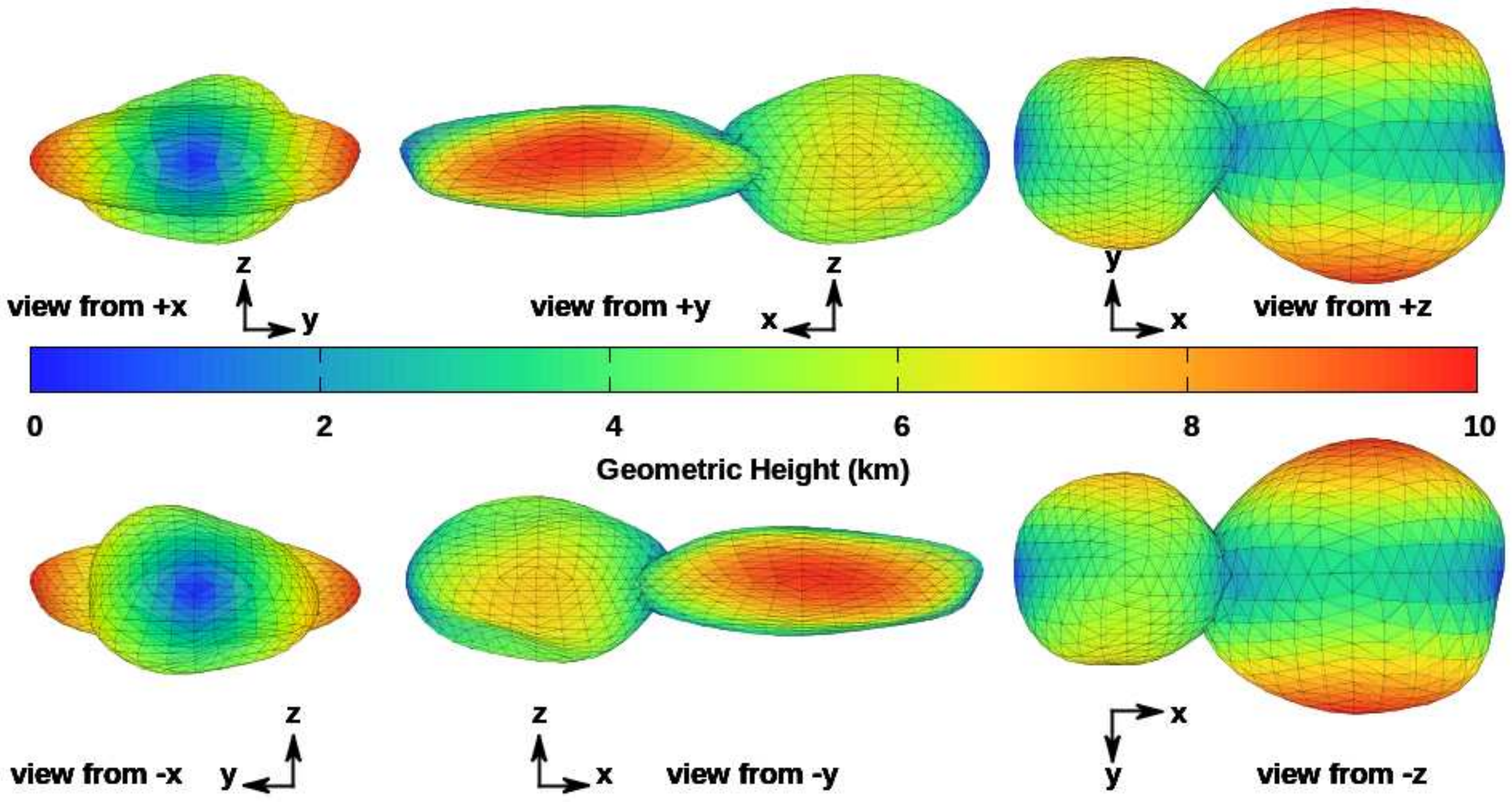}}
  \caption{Geometric polyhedral shape model in 3-D of Arrokoth shown in $6$ perspective views ($\pm x$, $\pm y$, and $\pm z$). The shape was built with $1,046$ vertices, $2,928$ edges, and $1,952$ triangular faces. The color code gives the centroid facet distance from the major axis $x$ in km (Geometric Height). An animated movie of our 3-D Arrokoth polyhedral shape model is available online (Movie 1).}
  \label{fig:prop_2}
\end{figure*}
\begin{figure*}
  \centering
  \fbox{\includegraphics[width=18cm]{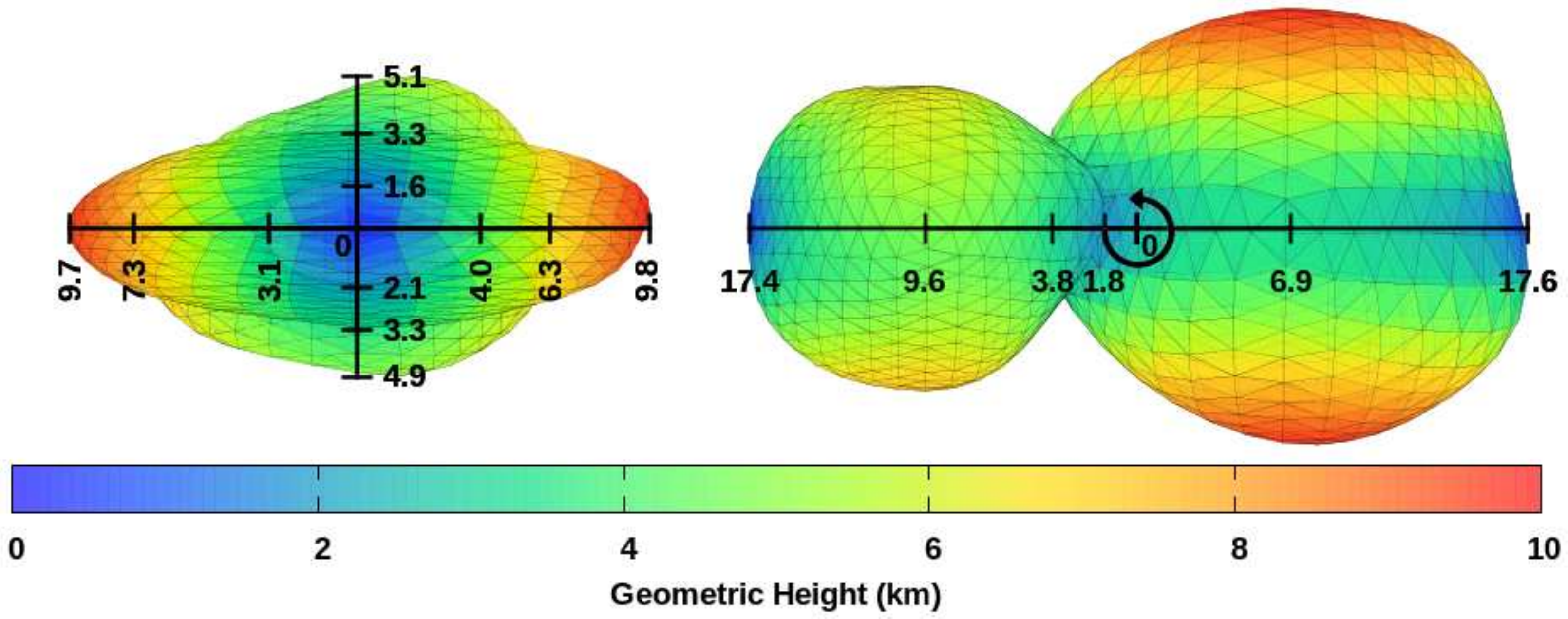}}
  \caption{(left-hand side) Arrokoth topographic shape viewed from the $+x$ major axis. The black bars show approximated width and length of the ``neck'', small and large lobes, respectively, measured from $x$-axis (0) in km. (right-hand side) Arrokoth shape viewed from the $+z$ major axis. The black bars indicate the distances from the Arrokoth geometric centre (0) along the $x$-axis, in km. The angular velocity vector is taken along the $z$-axis direction according to the right-hand rule and it also lies at the body centroid (0). The color code gives the Geometric Height in km.}
  \label{fig:prop_3}
\end{figure*}
An overview of the initial results from the New Horizons spacecraft's close-approach reconnaissance showed that Arrokoth has a peculiar shape. It is a contact binary with an overall major axis length of $31.7 \pm 0.5$\,km that is composed of two grossly spherical parental lobes with radii of $\sim 9.75$ and $\sim 7.1$\ km for the large lobe `Ultima' and the small lobe `Thule', respectively \citep{Stern2019}. This odd shape resembled a `snowman or bowling pin'. However, the initial results from New Horizons' space probe exploration show that the best-fitting shape for the Arrokoth contact binary planetesimal is approximately a lenticular shape with overall dimensions of $\sim 35 \pm 1 \times 20 \pm 1 \times 10 \pm 3$\,km, where the large and small lobes rotate around a common centre mass with a slow rotation period of $15.92 \pm 0.02$ h \citep{Stern2019b}.

In this work, we used the 3-D shape of Arrokoth based on the initial results of flyby images obtained with time observation using the New Horizons spacecraft's LORRI imager component during the close approach \citep{Stern2019b}.\footnote{We obtained polyhedral data for Arrokoth from the 3D Asteroid Catalogue website: \url{https://3d-asteroids.space/asteroids/486958-Arrokoth}.} For our purposes, we are only interested in the geometric vertices (v) of \citet{Stern2019b}'s best-fitting shape model. In Fig. \ref{fig:prop_2}, we reproduced the Arrokoth polyhedral model using $1,046$ vertices and $2,928$ edges combined into $1,952$ triangular faces.\footnote{We used the gnuplot program \citep{Williams2011} to build the triangular mesh of this figure.} We constructed our polyhedral model of Arrokoth using the same volume and overall dimensions computed from \citet{Stern2019b}. Our polyhedral model also has an approximate total volume of $2,428$\,km$^3$ and an `equivalent' spherical diameter of $\sim 8.3$\,km. The large lobe has a volume of $1,364$\,km$^3$, while the small lobe has a volume of $1,000$\,km$^3$, with each lobe having spherical diameters of $\sim 6.9$ and $\sim 6.2$, respectively. The thin `neck' of Arrokoth has a volume of only $64$\,km$^3$ with a spherical diameter of $\sim 2.5$\,km. Our measurements are also consistent with the estimated volumes obtained by \citet{Stern2019b}. Figure \ref{fig:prop_2} illustrates Arrokoth's 3-D polyhedral shape from $6$ perspectives: $\pm x$, $\pm y$ and $\pm z$. The colour box code gives the barycentre distance of each triangular face from the major axis $x$. The colours range from blue to red and highlight the large and small lobe shapes that resemble a `pancake' and a `walnut', respectively. Fig. \ref{fig:prop_2} shows that Arrokoth's shape is irregular, asymmetric, and non-convex. The projections of the shape in the equatorial planes $xOy$, $xOz$ and $yOz$ are totally different.
\begin{figure}
  \centering
  \fbox{\includegraphics[width=8.44cm]{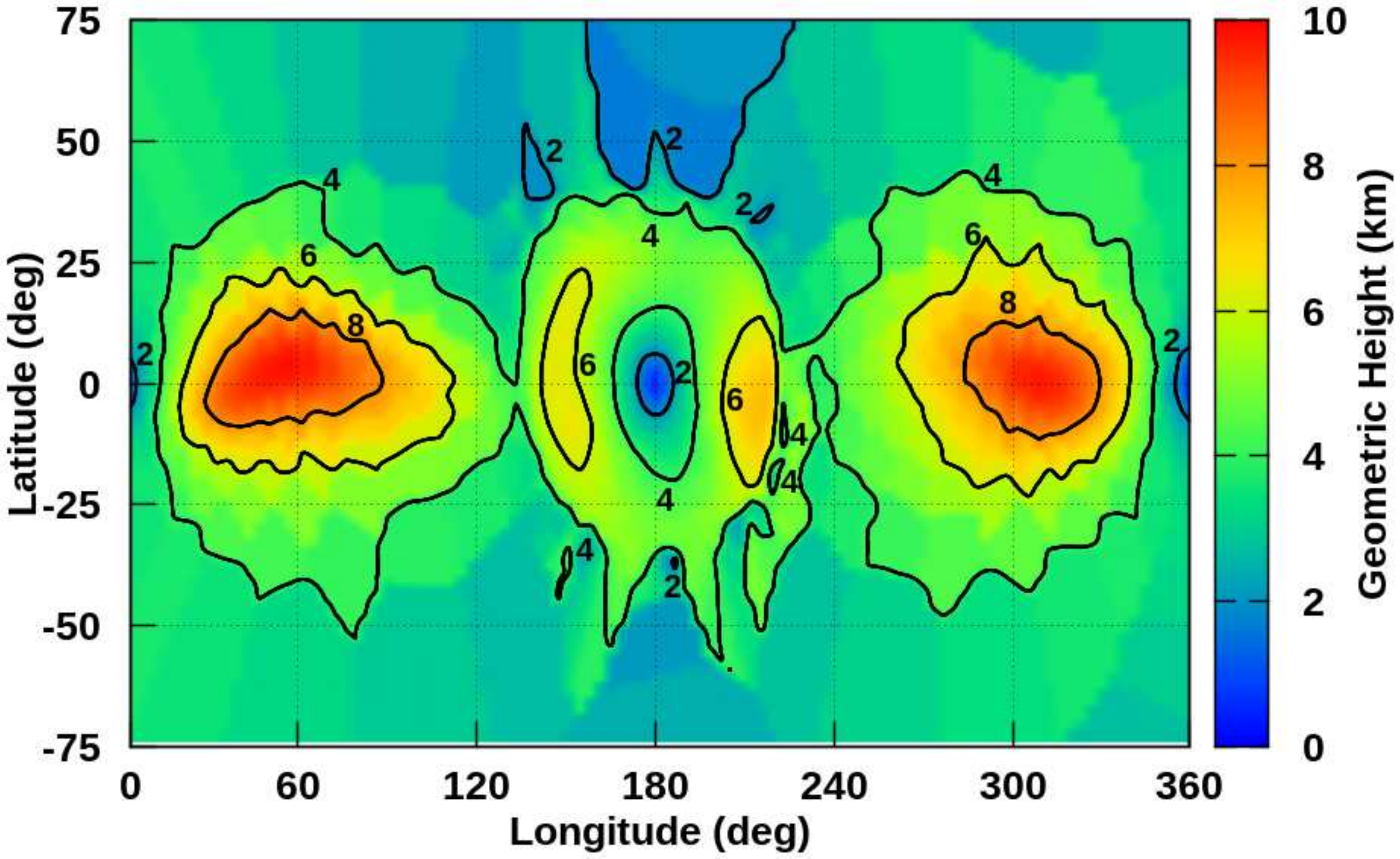}}
  \caption{Level curves contours of Arrokoth topographic. The color bar denotes the depth to the major axis $x$ in km (Geometric Height). The black contour lines represent levels every $2$ km.}
  \label{fig:prop_4}
\end{figure}
\begin{figure}
  \centering
  \fbox{\includegraphics[width=8.2cm]{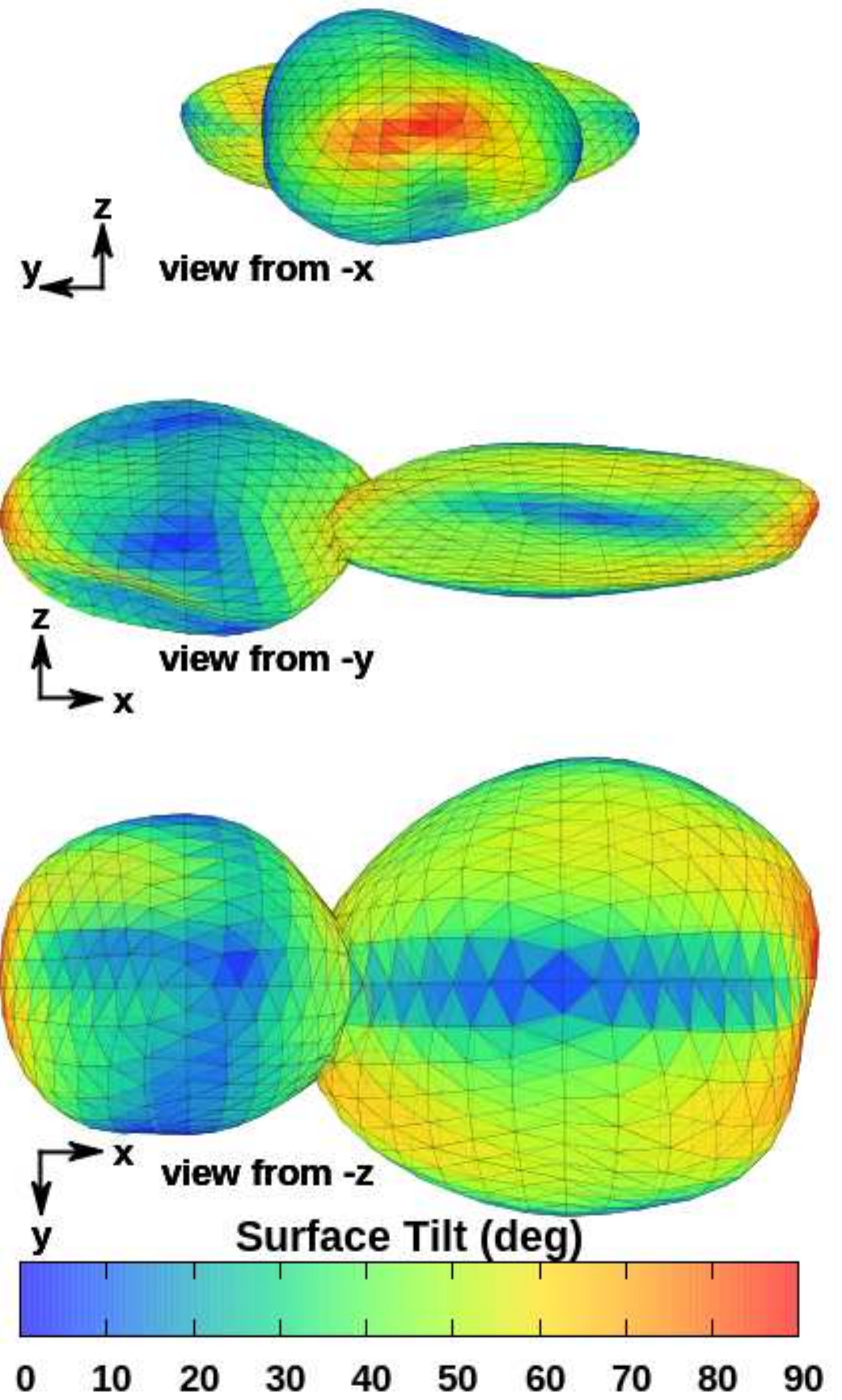}}
  \caption{Map of surface tilt angles computed across the surface of Arrokoth shown in perspective views $-x$, $-y$ and $-z$. The color code gives the angle (degrees) between the normal vector of each facet and the vector with the origin at the $x-$axis and ending at the face centroid.}
  \label{fig:prop_5}
\end{figure}

\subsection{Geometric Height}
\label{sec:prop:geoh}
We define the surface distance of each triangular face centroid from the major axis $x$ as \textit{geometric height}. This purely geometric quantity in addition to the surface tilt angles is useful when dealing with an elongated body topography where a descending spacecraft must land on the surface of an object. The geometric height could provide information about the $x-$axis' axial distance from the body surface to the space probe (e.g., see \citet{Chanut2015b}). In Fig. \ref{fig:prop_3}, we plotted Arrokoth's topographic shape from $+x$ (left side) and $+z$ (right side) axes. In the left-hand side of Fig. \ref{fig:prop_3}, the approximate width and length measures of Arrokoth's neck and its small and large lobes are shown in kilometres. From this perspective, the large lobe clearly has a lenticular shape, while the shape of the small lobe is more equidimensional. Our model of Arrokoth's shape fits into the bounding box of $-17.377359 \leq x \leq 17.563710$, $-9.734253 \leq y \leq 9.826064$ and $-4.931453 \leq z \leq 5.121795$\,km. Our model shows that the large lobe has dimensions of approximately $21.36 \times 19.56 \times 6.61$\,km, whereas the small lobe has dimensions of approximately $15.50 \times 13.73 \times 10.05$\,km, which are also consistent with the estimated uncertainties observed by \citet{Stern2019b}. The neck is between $3.8-1.8$\,km away from the left side of Arrokoth along the $x$-axis, and their $y$ and $z$ dimensions have dimensions of approximately $11.16 \times 6.22$\,km. The black lines on the right-hand side of Fig. \ref{fig:prop_3} shows the distances measured from the body's barycentre along the $x$-axis in kilometres. This figure shows that the centre mass of Arrokoth contact binary lies inside the large lobe and it is an offset $1.8$\,km from Arrokoth's neck. The edges of the large and small lobes are approximately equidistant from the barycentre. The geometric centres of the large and small lobes are $16.5$\,km apart.

A contour plot of Arrokoth's level curves is shown in Fig. \ref{fig:prop_4}. This contour plot used conversion of simple Cartesian co-ordinates to spherical co-ordinates, where the Cartesian co-ordinates are the vertices of each polyhedral triangular facet with respect to the centre mass. The colours highlight the geometric height. Both lobes have local maximum values for geometric height near the equator; thus, we can interpret this topographic feature as a mountain summit.

\subsection{Surface Tilts}
\label{sec:prop:tilt}
\textit{Surface tilt} angles are used to map the orientation of the surface of a body relative to a vector, which is usually measured from the centre mass of the body to the surface face centroid \citep{Scheeres2016}. Here, we used a different approach: i.e., we chose to use the geometric height vector to deal with this geometric feature. The geometric height vector is measured from the major axis $x$ distance point to the face's barycentre. We then used the dot product between this vector and the normal face vector to find the surface tilt angle of the local face. Figure \ref{fig:prop_5} shows our results. As expected, the largest value of the surface tilt angles did not exceed $90^\circ$, except for some cases at the edges of the lobes, with angles of up to $140^\circ$. Comparing perspectives of the large lobe, i.e., $-x$, $-y$ and $-z$, we can conclude that between the equator and the poles, the surface tilts have the most highest values, with most between $40^\circ-60^\circ$. However, the values decrease to close to zero at the poles, where the surface tilts are mostly longitudinally uniform. Meanwhile, the small lobe shows a type of longitudinal non-uniformity on its surface tilts, which is caused by the lobe's asymmetrical shape. This finding implies the existence of some craters on the small lobe's surface \citep{Spencer2020}. Arrokoth's neck has most of the surface tilts around $60^\circ$ in addition to most of the highest Arrokoth surface tilt angles (except at body edges) as can be seen from perspective $-y$. This same analysis can be extended for the other projection planes.

Arrokoth's mass remains unknown because no satellites could better estimate its mass \citep{Stern2019,Stern2019b}. Its density is also poorly constrained and the large and small lobes are expected to have the same density as their volume ratio of $\sim 1.36$. Arrokoth's surface colour suggests that the presence of less water ice on Arrokoth than on Nix, a satellite of Pluto similar in size to Arrokoth \citep{Stern2019,Grundy2020}. To estimate Arrokoth's mass for our model, we assume a uniform mean bulk density of $\rho=0.5$\,g\,cm$^{-3}$ derived from cometary nuclei models \citep{Stern2019b} that give us a total body mass of approximately $M=1.2138 \times 10^{15}$\,kg. The masses of the large and small lobes are $6.819 \times 10^{14}$ and $5.000 \times 10^{14}$\,kg, respectively. The neck has a mass of $3.190 \times 10^{13}$\,kg. In addition, we compute Arrokoth's principal moments of inertia as normalized by the total body mass\footnote{A modified version of the \citet{Mirtich1996} algorithm was used to compute Arrokoth's principal moments of inertia, as well as those for each individual lobe.}:
\begin{align}
J_{xx}/M & = 18.279667  \,\text{km}^2, \nonumber \\  
J_{yy}/M & = 89.744625  \,\text{km}^2, \nonumber \\  
J_{zz}/M & = 101.682581 \,\text{km}^2. \nonumber   
\end{align}
Table \ref{tab:prop_4} shows the values for Arrokoth's principal moments of inertia as normalized by the respective lobe mass $M_l$ ($l=1,2$) for the large and small lobes.
\begin{table}
 \centering
  \caption{Values of the principal moments of inertia, (normalized by the respective mass of each body), the gravitational coefficients $C_{20}$ and $C_{22}$ (normalized by the respective mass and squared spherical radius of each body), and of mass-distribution dimensionless parameter $\sigma$ calculated for each Arrokoth lobe. $M_l$ ($l=1,2$) represents the mass for the large (1) and small (2) lobes, respectively.}
 \label{tab:prop_4}
 \begin{tabular}{ccc}
  \toprule
        & Large & Small \\
  \hline
  $J_{xx}/M_l$ (\text{km}$^2$) & 15.723384 & 11.951184 \\
  $J_{yy}/M_l$ (\text{km}$^2$) & 22.016971 & 15.384692 \\
  $J_{zz}/M_l$ (\text{km}$^2$) & 33.283804 & 17.246621 \\
  \hline
  $C_{20}$ & -0.309325 & -0.097417 \\
  $C_{22}$ & 0.033766  & 0.023366 \\
  \hline
  $\sigma$ &  0.36  &  0.65 \\
  \hline
 \end{tabular}
 \end{table}

\section{Mathematical Model}
\label{sec:grav}
The New Horizons spacecraft's flyby images suggest that the shape of Arrokoth's minor body can be considered as two roughly spherical but flattened lobes brought into contact (Fig. \ref{fig:prop_3}). For our purposes, therefore, we consider that Arrokoth's gravitational field is generated by a near-contact polyhedral binary system. This approach has been widely used to study the nearby dynamic environments around binary small bodies (e.g., see \citet{Scheeres2006,Fahnestock2008,Bellerose2008,McMahon2010,Yu2013,Feng2016,Yu2017,Jiang2018,Lan2018,Shi2018,Zeng2018}). 
For example, \citet{Feng2016} use a shape model of a contact binary body consisting of ellipsoidal and spherical components in physical contact to study the asteroidal contact binary system (1996) HW1. Our non-convex polyhedral model was adopted to consider the irregularity of binary gravitational fields that represent irregular shapes and the corresponding gravitational fields of the primary (large) and secondary (small) lobes. Binary minor bodies are relatively common; therefore, our methods will be applicable to other similar binary systems with polyhedral model data. Our results can provide general insights into the dynamic environments and orbital behaviours in the vicinity of these binary bodies.

\subsection{Elongation and Oblateness}
\label{sec:prop:elobl}
We can also compute a reference triaxial ellipsoid for Arrokoth and each lobe using their principal moments of inertia. According to \citet{Dobrovolskis1996}, every diagonal inertia tensor is identical to that of an equivalent ellipsoid of mass $M$ and principal semi-axes $a \geq b \geq c$, with corresponding principal moments $J_{xx}$, $J_{yy}$ and $J_{zz}$. The principal semimajor axes $a$, $b$ and $c$ are solved as follows:
\begin{eqnarray}
a=\sqrt{\frac{5(J_{yy}+J_{zz}-J_{xx})}{2M}}, \nonumber \\
b=\sqrt{\frac{5(J_{xx}+J_{zz}-J_{yy})}{2M}}, \\
c=\sqrt{\frac{5(J_{xx}+J_{yy}-J_{zz})}{2M}}. \nonumber
\label{eq:prop_4}
\end{eqnarray}

This configuration leads us to consider the Arrokoth lobes as triaxial ellipsoids with overall dimensions in each principal semi-axes $a$, $b$ and $c$ given in Table \ref{tab:prop_1}.
\begin{table}
 \centering
  \caption{Overall semi-axes dimensions (km) of the Arrokoth lobes.}
 \label{tab:prop_1}
 \begin{tabular}{cccc}
  \hline
  Lobe & a & b & c \\
  \hline
  Large & 9.9 & 8.2 & 3.3 \\
  Small & 7.2 & 5.9 & 5.0 \\
  \hline
  Arrokoth & 20.8 & 8.7 & 4.0 \\
  \hline
 \end{tabular}
 \end{table}
For the Arrokoth contact binary, our model used Eq. \eqref{eq:prop_4} to provide a body with principal semi-axes of $20.8 \times 8.7 \times 4.0$ km, which are close to the overall dimensions of the body.

From the moments of inertia, we computed the two main terms $C_{20}$ ($-J_2$) and $C_{22}$ of the harmonic expansion that corresponds to the second-order and the degree of gravity harmonic coefficients and expressed the irregular shape of the mass distribution of a body \citep{MacMillan1958}. Our model yielded the following values for gravitational field terms (divided by the body mass and squared spherical radius):
\begin{align}
C_{20} & = -\frac{1}{2MR_s^2}(2J_{zz}-J_{xx}-J_{yy}) = -0.685765 , \nonumber \\  
C_{22} & =  \frac{1}{4MR_s^2}(J_{yy}-J_{xx}) = 0.257016.
\label{eq:prop_5}
\end{align}

If we choose a measure of $R_s=8.337$\,km (equivalent spherical radius) for the normalization radius and knowing that $J_2=-C_{20}$, then we get from Eq. \eqref{eq:prop_5} the dimensionless values of $J_2=-6.858\times 10^{-1}$ and $C_{22}=2.570\times 10^{-1}$ for Arrokoth's gravitational field coefficients. The zonal coefficient $J_2$ gives an idea of Arrokoth's equatorial \textit{oblateness} and the tesseral coefficient $C_{22}$ gives an idea of the equatorial deformation due to the mutual interaction of both lobes. The zonal $J_2$ and tesseral $C_{22}$ coefficients are of the same order of magnitude for Arrokoth, which means that Arrokoth has a highly oblate as well as elongated shape. We computed the zonal and tesseral gravitational coefficients for each lobe and the results are shown in Table \ref{tab:prop_4}. As expected from Fig. \ref{fig:prop_3}, the large lobe has higher oblateness as well as \textit{elongation} than the small lobe. The large lobe had an oblateness $\sim \,1.72\times$ greater than that observed by \citet{Grishin2020}. The small lobe had an approximately equal oblateness of $0.1$ to that observed by \citet{Grishin2020}. These authors used ellipsoidal models for the large and small lobes.

\begin{figure}
  \centering
  \includegraphics[width=8.6cm]{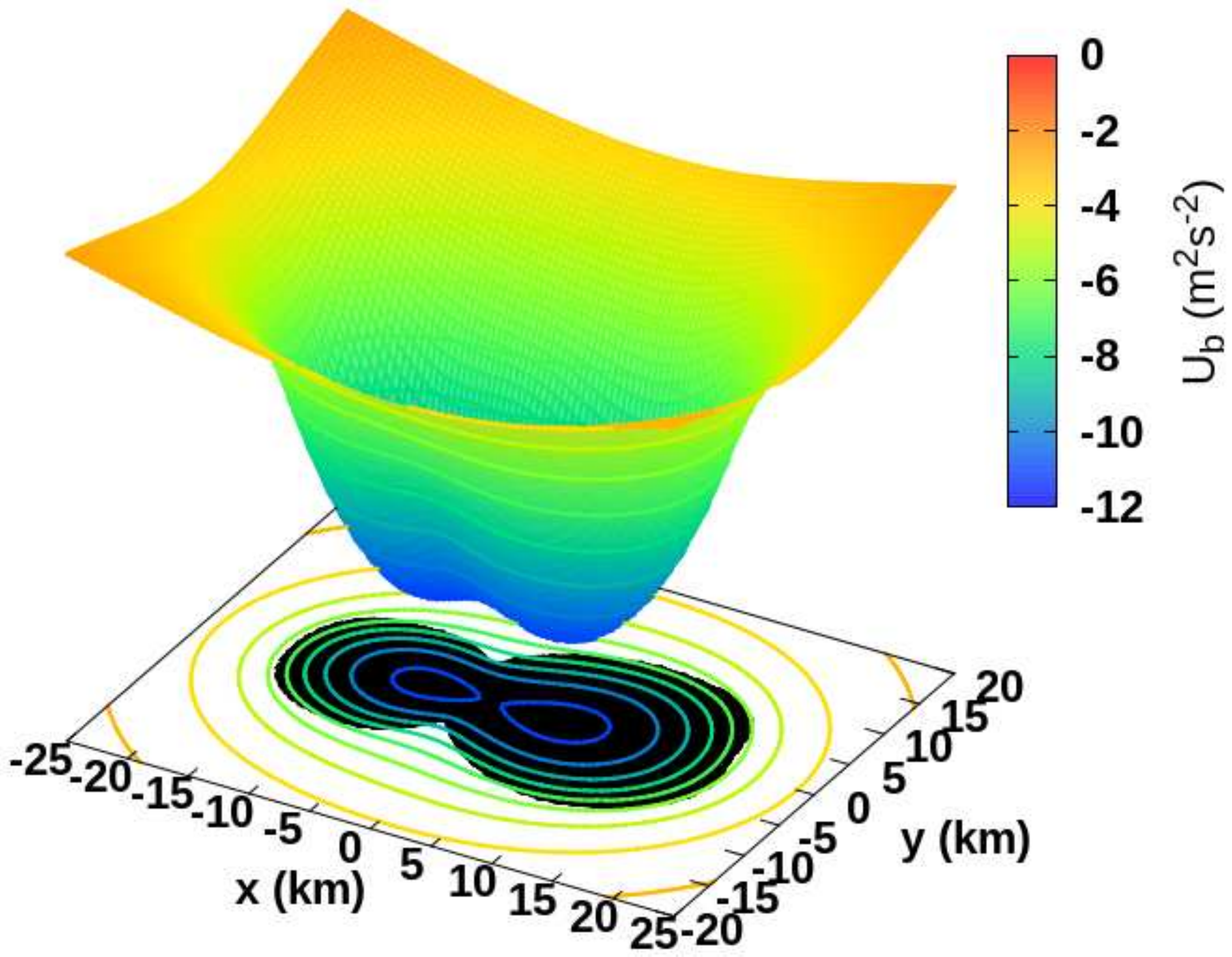}\\
  \vspace{1.05cm}
  \includegraphics[width=8.6cm]{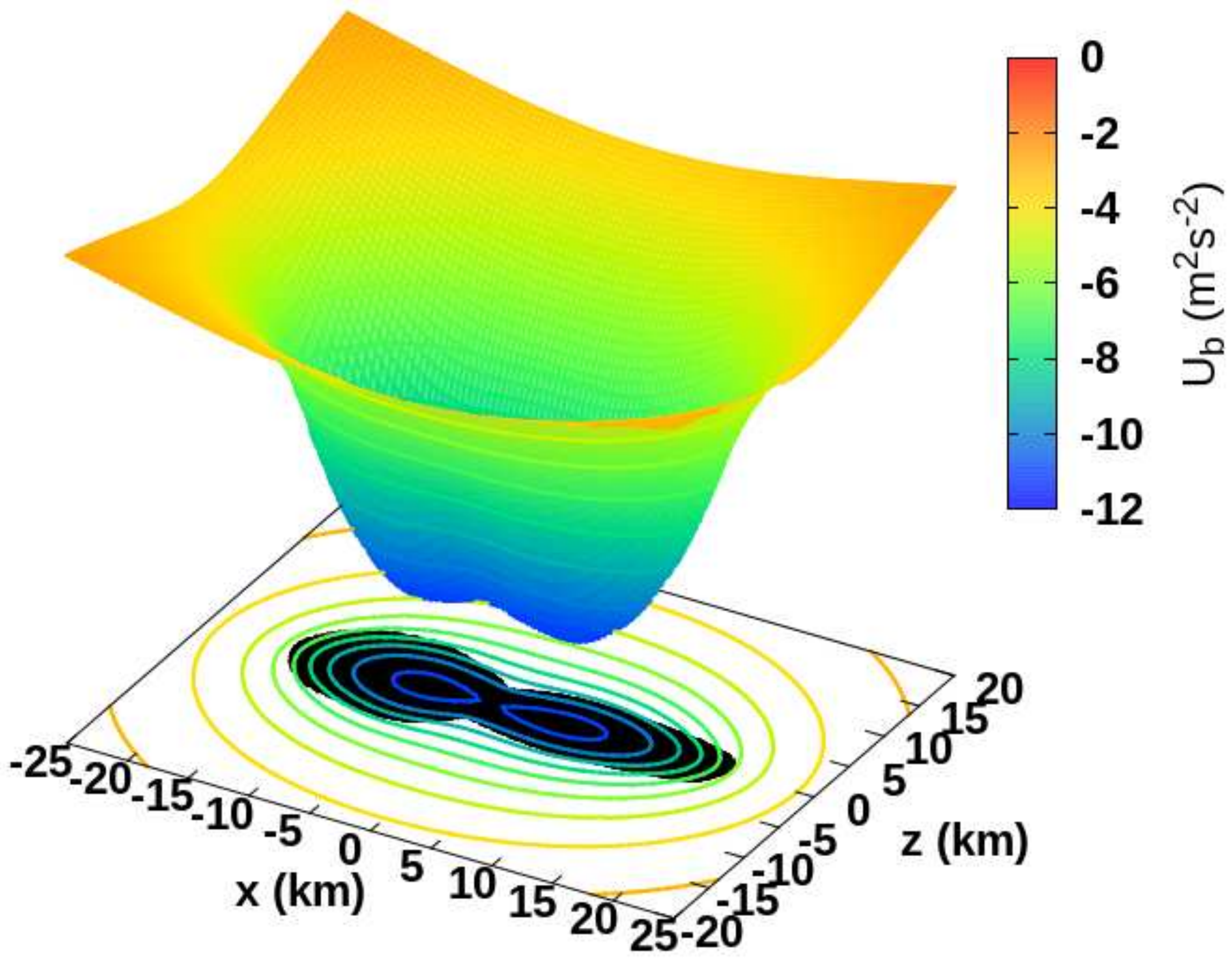}\\
  \vspace{1.05cm}
  \includegraphics[width=8.6cm]{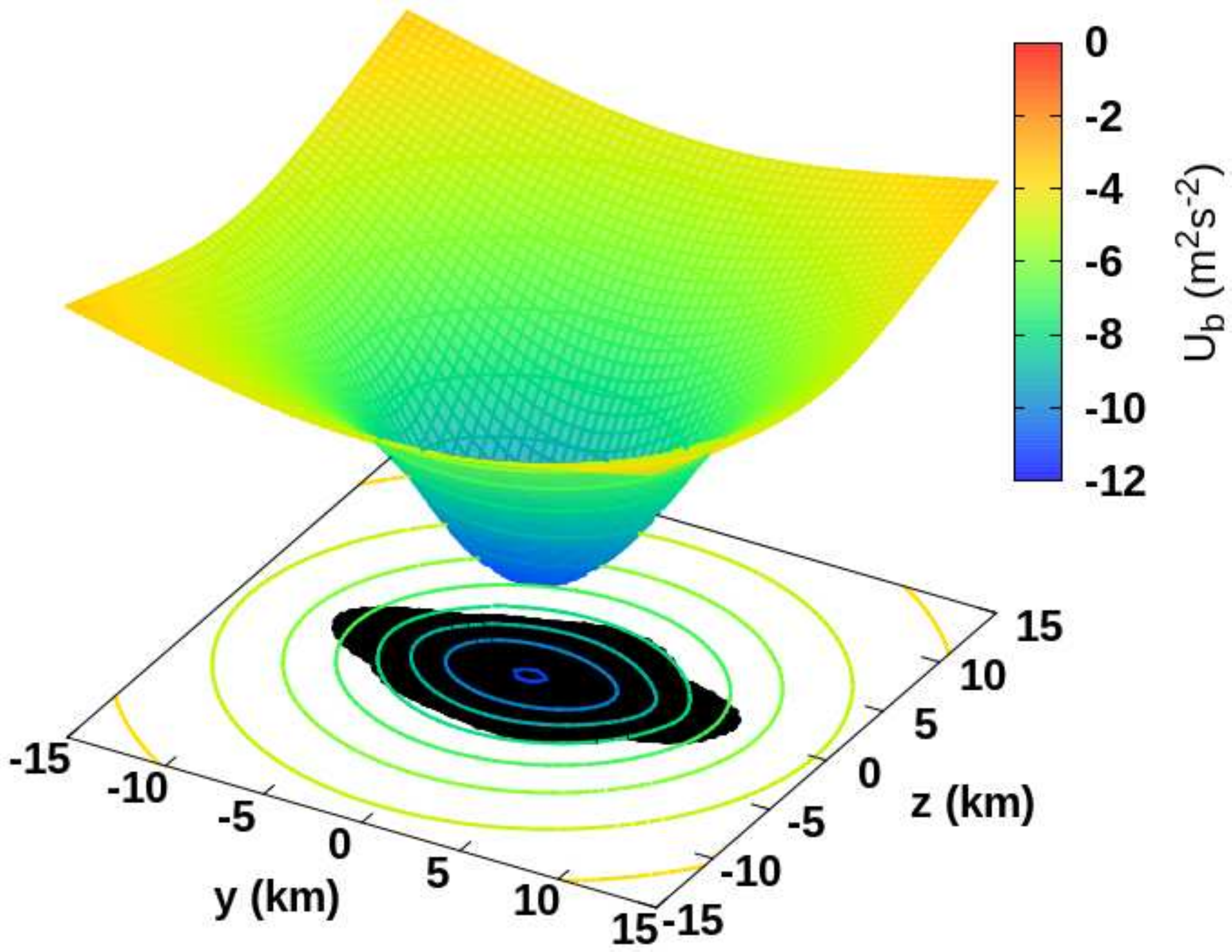}
  \caption{Numerical gravitational force potential energy of Arrokoth contact binary in the $xOy$, $xOz$, and $yOz$ planes, respectively. The contour lines represent the binary equipotential curves and the color code gives the intensity of the Arrokoth mutual gravitational force potential $U_b$ in m$^2$\,s$^{-2}$.}
  \label{fig:grav_2}
\end{figure}

We applied \citet{Werner1997}'s numerical algorithm in the polyhedral model of Arrokoth considering a uniform bulk density to calculate the gravitational spherical harmonic coefficients $C_{n,m}$, and $S_{n,m}$ up to order and degree $4^{th}$. Table \ref{tab:prop_3} summarizes our numerical results for Arrokoth's gravitational field coefficients. In contrast to the gravitational field coefficients presented in the literature, we find it more useful to present the normalized coefficients in Table \ref{tab:prop_3}. These numerical gravity coefficients must be fully normalized up to order and degree $10^{th}$ onwards to avoid divergence because of the order of their magnitudes according to the formulae presented in \citet{MacMillan1936} and \citet{Kaula1966}. Table \ref{tab:prop_3} shows that the zonal gravity coefficients $C_{20}$ ($-J_2$) and $C_{40}$ ($-J_4$) have a closer order of magnitude in absolute values than zonal gravity $C_{30}$ ($-J_3$). These results reveal an irregular gravitational field with a tear-drop shape pointing towards the small lobe (Fig. \ref{fig:grav_2}) that differs, e.g., from the Earth's gravitational field, which is more spherical \citep{Pavlis2008}.

\citet{Hu2004} define a mass-distribution dimensionless parameter $\sigma$ to study the measurement of a body's shape using its gravitational field.
\begin{align}
\sigma & = \frac{J_{yy}-J_{xx}}{J_{zz}-J_{xx}} = -\frac{4C_{22}}{C_{20}-2C_{22}} = 0.86,
\label{eq:prop_6}
\end{align}
\noindent where $0 \leq \sigma \leq 1$ for any mass distribution with $J_{xx} \leq J_{yy} \leq J_{zz}$. If $\sigma=0$, the body has a rotational symmetry about the $z$-axis ($J_{yy} = J_{xx}$), while one with a value of $\sigma=1$ denotes a body with rotational symmetry about the $x$-axis ($J_{yy} = J_{zz}$). This value of $\sigma = 0.86$ from Eq. \eqref{eq:prop_6} leads us to conclude that Arrokoth has a near rotational symmetry about the $x$-axis, i.e., it has an equivalent highly prolate shape (Jacobi ellipsoid) like asteroids (433) Eros, (216) Kleopatra \citep{Chanut2014,Chanut2015b} and the inferred shape of interstellar object 1I/2017 U1 `Oumuamua' \citep{Meech2017,Katz2018,Hui2019,Vazan2020,Zhang2020}. The dimensionless mass distribution parameters $\sigma$ for the large and small lobes are also given in Table \ref{tab:prop_4}. The results show that the large lobe ($\sigma = 0.36$) has a close rotational symmetry about the $z$-axis like asteroid (101955) Bennu \citep{Amarante2019}. The small lobe ($\sigma = 0.65$) denotes a parameter value that gives a rotational symmetry between $z$-axis ($\sigma = 0$) and $x$-axis ($\sigma = 1$), as asteroid (21) Lutetia \citep{Aljbaae2017}.
\begin{table}
 \centering
  \caption{Arrokoth normalized numerical gravity field harmonics coefficients up to order and degree $4^{th}$, using the polyhedral shape model. These coefficients are computed using the reference radius distance of $R_s=8.337$ km. The frame is located at the centre of mass and aligned with the principal moments of inertia.}
 \label{tab:prop_3}
 \begin{tabular}{cccc}
  \hline
  Order (n) & Degree (m) & $C_{n,m}$ & $S_{n,m}$ \\
  \hline
  0 & 0 & 1.0 & - \\
  1 & 0 & 0 & - \\
  1 & 1 & 0 & 0 \\
  2 & 0 & $-6.857653\times 10^{-1}$ & - \\
  2 & 1 & 0 & 0 \\
  2 & 2 & $2.570156 \times 10^{-1}$ & 0 \\
  3 & 0 & $3.973648 \times 10^{-3}$ & - \\
  3 & 1 & $-1.765315\times 10^{-2}$ & $1.904884 \times 10^{-3}$ \\
  3 & 2 & $-5.392922\times 10^{-4}$ & $-1.610207\times 10^{-3}$ \\
  3 & 3 & $-1.190173\times 10^{-2}$ & $-3.995327\times 10^{-4}$ \\
  4 & 0 & $1.030572$ & - \\
  4 & 1 & $2.667503 \times 10^{-3}$ & $3.726019 \times 10^{-3}$ \\
  4 & 2 & $-9.218806\times 10^{-2}$ & $-5.937414\times 10^{-4}$ \\
  4 & 3 & $-4.096281\times 10^{-4}$ & $-6.703860\times 10^{-4}$ \\
  4 & 4 & $6.406631 \times 10^{-3}$ & $-3.255937\times 10^{-4}$ \\
  \hline
 \end{tabular}
 \end{table}

\subsection{Binary Gravitational Force Potential}
\label{sec:grav:pot}
The formulation of the \textit{binary gravitational force potential} $U_b$ used herein is based on representing each lobe as a closed polyhedron with triangular faces. We also considered that Arrokoth is a complete homogeneous contact binary body with constant density. We used the polyhedral approach to compute the mutual gravitational force potential of a polyhedron. Based on these assumptions, the binary gravitational force potential $U_b$ can be computed using Equation \eqref{eq:grav_0} by the sum of individual ones $U_1$ and $U_2$ of each lobe, respectively:
\begin{eqnarray}
U_b(x,y,z) = U_1(x,y,z)+U_2(x,y,z),
\label{eq:grav_0}
\end{eqnarray}
\noindent where $x$, $y$ and $z$ represent the co-ordinates of a massless particle in the binary body-fixed frame measured from the barycentre of the binary system, with the unit vectors defined along the minimum, intermediate and maximum moments of inertia, respectively.

The models developed by \citet{Werner1994}, \citet{Petrovic1996} and \citet{Werner1997b} provide a particularly convenient and robust analytic solution in the calculation of the gravitational force potential and its derivatives due to a homogeneous polyhedron with polygonal faces. We briefly describe the numerical polyhedra method used in our work \citep{Petrovic1996,Tsoulis2001} in Appendix \ref{sec:app}.

\begin{figure}
  \centering
  \includegraphics[width=8.6cm]{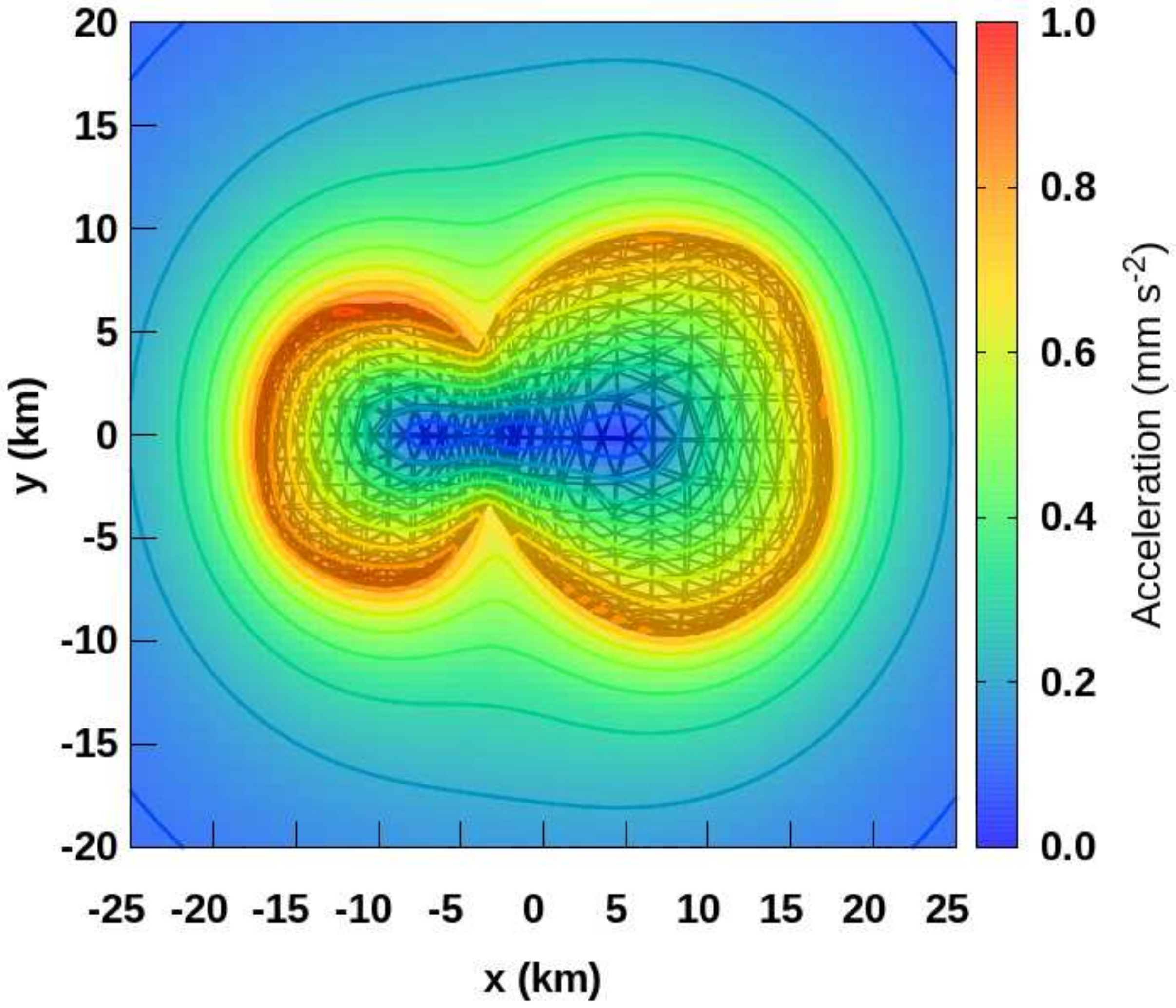}\\
  \includegraphics[width=8.6cm]{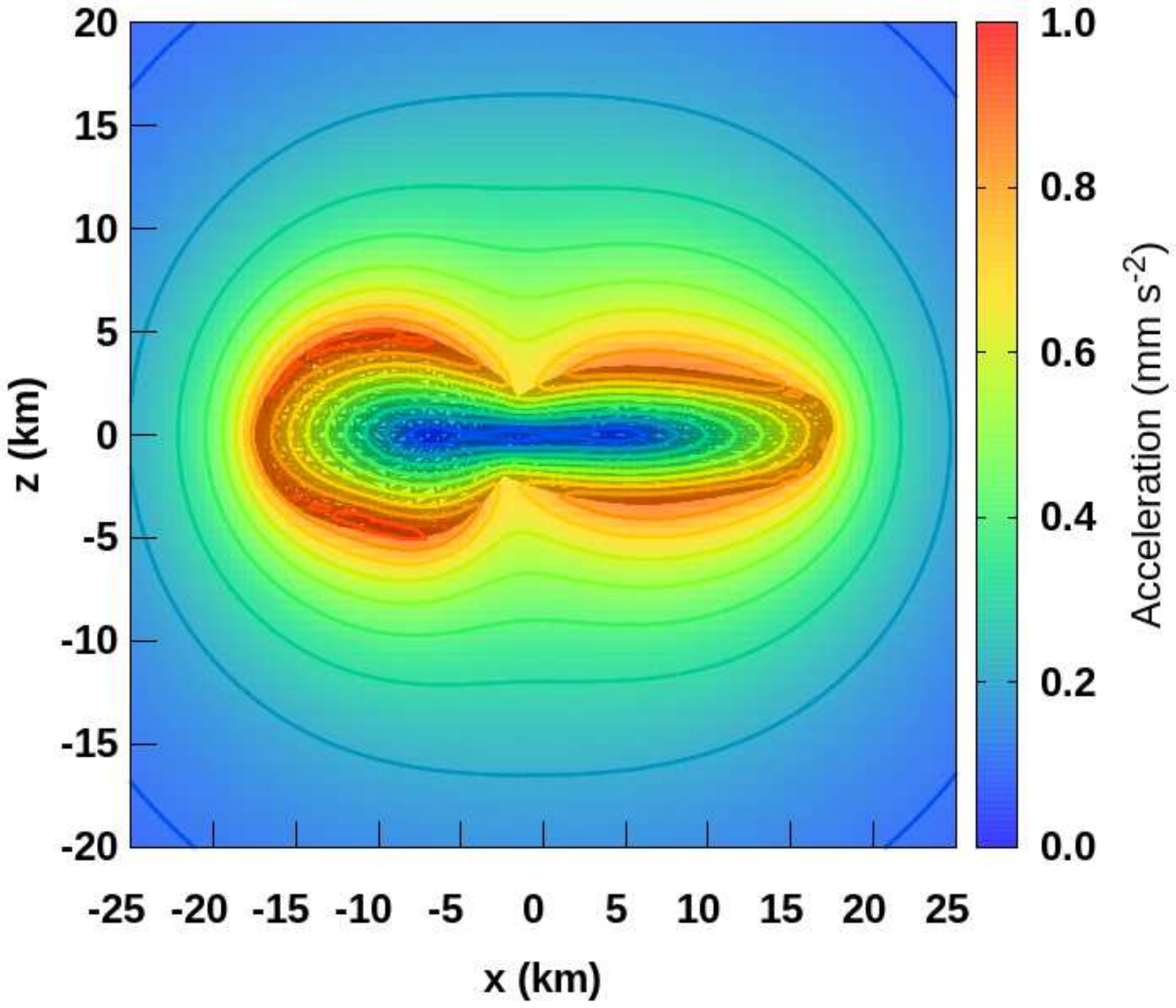}\\
  \includegraphics[width=8.6cm]{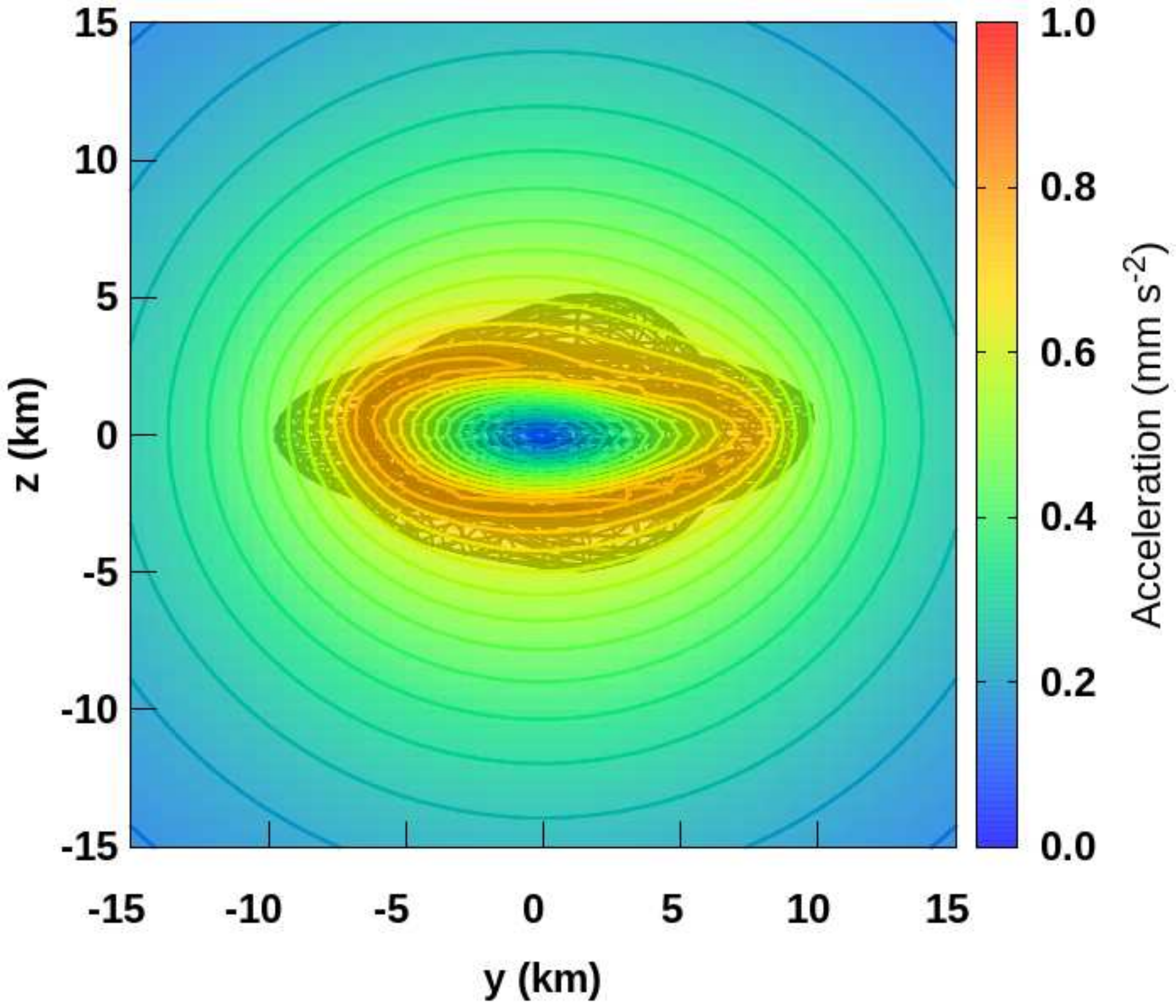}
  \caption{Numerical gravity field of Arrokoth contact binary in the $xOy$, $xOz$, and $yOz$ planes, respectively. The contour levels denote the lines of force and the color panel gives the intensity (acceleration) of the gravity attraction vector $|-\nabla U_b|$ in mm\,s$^{-2}$.}
  \label{fig:grav_3}
\end{figure} 

\subsection{Binary Equipotential Curves}
\label{sec:grav:eq}
Figure \ref{fig:grav_2} shows Arrokoth's potential energy surfaces, which were numerically computed using the polyhedral approach through the Minor-Gravity package from the $xOy$, $xOz$ and $yOz$ projection planes. From the potential energy surfaces, we plotted the binary equipotential curves in each projection plane. The contour lines denote the binary equipotential curves and the line colours show the Arrokoth binary gravitational force potential $U_b$ in m$^2$\,s$^{-2}$. The model of Arrokoth's model is also plotted in the projection planes (perspectives $+z$, $-y$ and $+x$, respectively) with black lines. Fig. \ref{fig:grav_2} shows that the binary force potential energy $U_b$ in the vicinity of Arrokoth lies approximately between (blue to green) a minimum value of $-10.51$ m$^2$\,s$^{-2}$ and a maximum value of $-6.27$ m$^2$\,s$^{-2}$. We can also show that the binary force potential has a global minimum value of $-11.30$ m$^2$\,s$^{-2}$, which has an approximated offset by $+4.5$\,km in the $x$-axis direction from Arrokoth's centre mass and it is inside the large lobe. Notably, the small lobe also has a local minima point slightly higher than the large lobe. The binary equipotential curves from projection planes $xOy$ and $xOz$ of Fig. \ref{fig:grav_2} are quite similar. From projection plane $xOz$, these curves seem more oblate than those that emerge from the $xOy$ projection plane. The reason for this result is in Arrokoth's contact binary shape as discussed in section \ref{sec:prop}. The shape of binary equipotential curves resembles a `teardrop' pointing towards the small lobe from inside the Arrokoth contact binary to its surface.

\subsection{Lines of Force}
\label{sec:grav:lin}
Figure \ref{fig:grav_3} shows the results of the acceleration over the surface and inside the Arrokoth contact binary. The contour maps denote the lines of force and the coloured levels show the intensity of the binary gravity attraction vector $|-\nabla U_b|$, in mm\,s$^{-2}$. Arrokoth contact binary's shape is also represented with a grey shadow in the perspectives $+z$, $-y$ and $+x$, respectively. We can observe from the $xOy$ and $xOz$ plots that the acceleration is slightly higher at the small (red) lobe boundary than the large (orange) lobe boundary with a maximum approximated local value of $\sim 9.64 \times 10^{-1}$ mm\,s$^{-2}$. In addition, the numerical surface acceleration computed across the neck is lower than the acceleration computed over the large and the small lobe boundaries with a minimum approximated local value of $\sim 5.89 \times 10^{-1}$ mm\,s$^{-2}$.

\section{Surface environment dynamics}
\label{sec:forc}
In this section, we present our detailed results for \textit{geopotential surface}, \textit{surface accelerations}, \textit{surface slopes} and \textit{escape speed} over Arrokoth contact binary's entire surface. Previous studies \citep{Spencer2020,McKinnon2020} have already presented some partial results for these features. In addition, we also show Arrokoth's surface stability through dynamic slope angles.

We define a \textit{binary geopotential} function in the following subsection to describe conservative quantities, as we shall see further on. We followed the same simple geopotential function definition as \citet{Scheeres2015}, which is sometimes also called a modified, effective, rotational, or pseudopotential function.

\subsection{Binary Geopotential}
\label{sec:grav:geo}
We specify the \textit{binary geopotential} $V_b(x,y,z)$ as a mathematical function of a position vector that combines the mutual gravitational force potential energy (Eq. \eqref{eq:grav_0}) from both lobes (index $1$ for the primary large lobe and index $2$ for the secondary small lobe) in addition to the effective contribution (centripetal potential) from the spin velocity vector $\pmb{\Omega}$ of the contact binary system. When computed over a body, the binary geopotential is a significant quantity that can be used to express the amount of energy flowing on the surface and within a bi-lobed body. It is directly related to the stress experienced internally by a spinning near-contact binary body \citep{Katz2019,Prentice2019,Stern2019b}. When combined with kinetic energy relative to the binary body-frame system, the binary geopotential results in a conserved quantity that reduces the dynamic motion of a particle in some allowed regions on the binary body-fixed frame. It can also be useful to compute the accelerations in the binary rotating frame that act on a particle given its location vector. In this section, some of their features over each Arrokoth lobe surface are computed as the binary geopotential surface, surface accelerations, surface slopes and surface escape speed. The expression of the binary geopotential in the binary rotating frame takes on a simpler form:
\begin{align}
V_b(x,y,z) &= -\frac{1}{2}\omega^2(x^2+y^2)+U_1(x,y,z)+U_2(x,y,z),
\label{eq:grav_1}
\end{align}
\noindent where the first right-hand side term is the centripetal potential with $|\pmb{\Omega}|=\omega$; $U_1(x,y,z)$ and $U_2(x,y,z)$ describe the gravitational force potential energy from the large and small lobes, respectively. Eq. \eqref{eq:grav_2} shows that their negative signal denotes attractive binary geopotential.
\begin{figure*}
  \centering
  \fbox{\includegraphics[width=\linewidth]{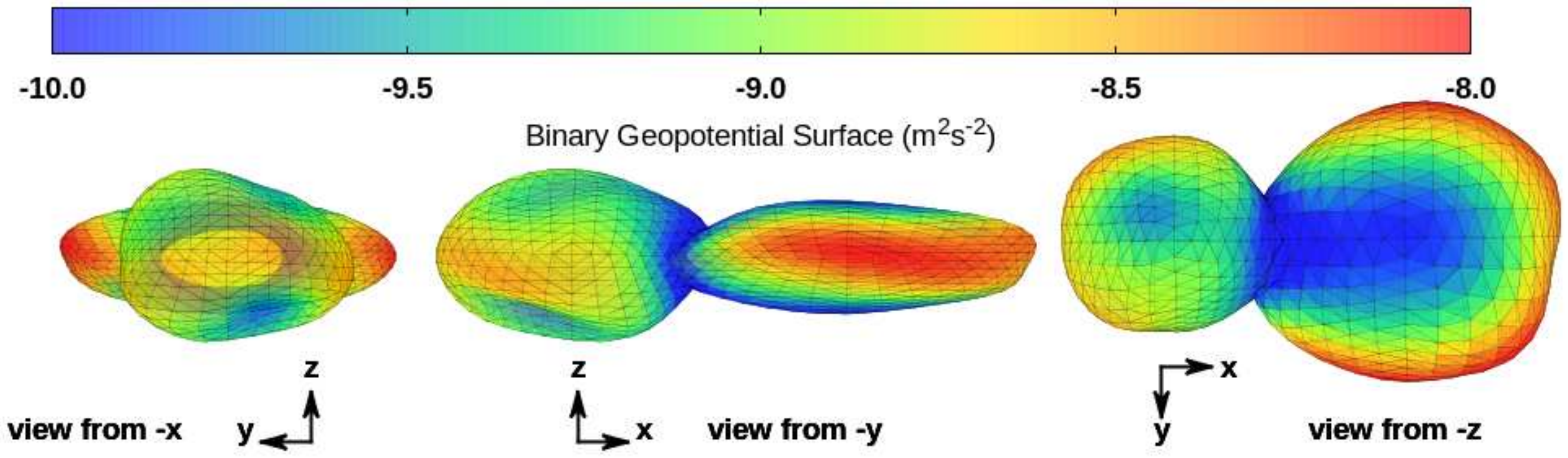}}
  \caption{Map of the binary geopotential computed across the surface of the Arrokoth contact binary. The color bar gives the numeric values of Equation \ref{eq:grav_1}, in m$^{2}$\,s$^{-2}$.}
  \label{fig:forc_1}
\end{figure*}
\begin{figure*}
  \centering
  \fbox{\includegraphics[width=\linewidth]{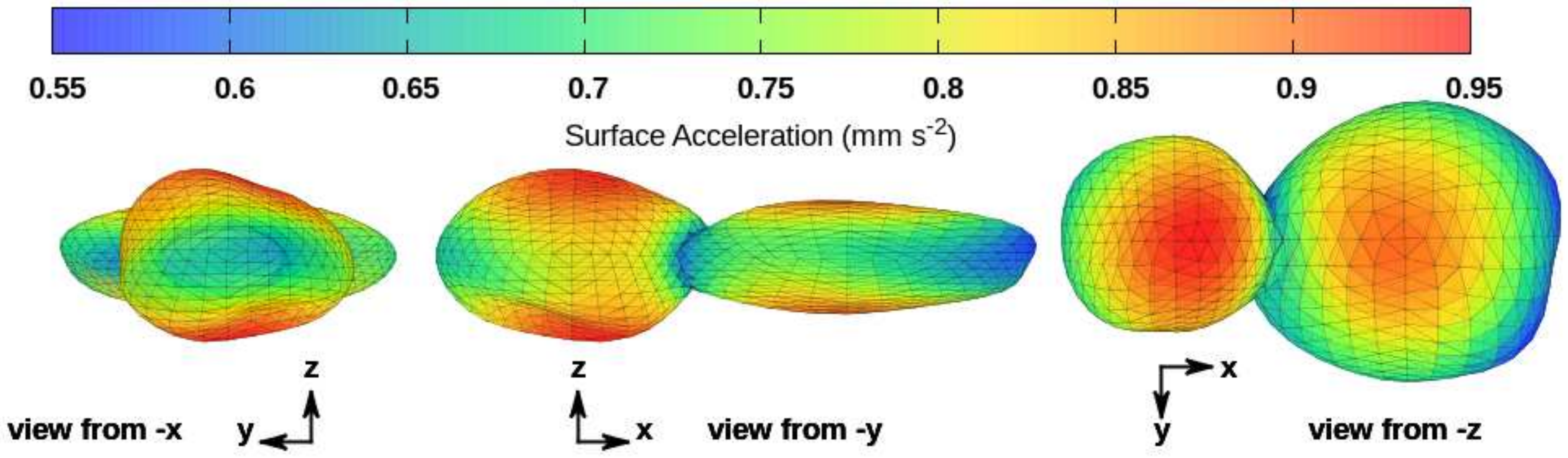}}
  \caption{Gravitational acceleration computed over neck, large and small lobe surfaces shown in $3$ perspective views ($-x$, $-y$ and $-z$). The color code gives the intensity of the total gravity attraction vector $|-\nabla V_b(x,y,z)|$ on the surface of Arrokoth contact binary, in mm\,s$^{-2}$.}
  \label{fig:forc_2}
\end{figure*}

\subsection{Binary Geopotential Surface}
\label{sec:forc:geo}
The equatorial regions of the large and small lobes suffer the influence of a maximum \textit{binary geopotential surface} with respect to the poles, which have the lowest values. Figure \ref{fig:forc_1} shows the binary geopotential computed across Arrokoth's surface. As can be seen in the large lobe, this influence is almost $1.3$ times higher at the equator than at the poles, while the small lobe has a binary geopotential surface local point of maxima value of $-8.38$ m$^{2}$\,s$^{-2}$ at the equatorial region and a local point of minima value of $-9.98$ m$^{2}$\,s$^{-2}$ at its poles. At the neck, this influence is further minimized ($-10.48$ m$^{2}$\,s$^{-2}$). These findings are in accordance with the geometric height feature shown in Fig. \ref{fig:prop_2}. As the binary gravitational force potential (Eq. \ref{eq:grav_2}) is attractive from Eq. \ref{eq:grav_1}, if a triangular facet centroid is farther away from the $x$-axis, i.e., it has a higher geometric height, then the binary geopotential surface should be higher at this location. This result explains why Arrokoth contact binary has higher binary geopotential surface values at the equatorial regions than at the poles. However, we do not consider the centripetal potential component of Eq. \ref{eq:grav_1}. For minor bodies with low spin periods, the centripetal potential influence interferes significantly in the geopotential effect across their surfaces. For example, the asteroid (101955) Bennu's geopotential is influenced by its rotational potential \citep{Scheeres2016}. However, for the Arrokoth contact binary, which has a high $15.92$\,h rotation period, its geopotential influence is not significant. Although the correct way to analyze this behaviour would be to consider the vector radius from Arrokoth's centre mass, the geometric height could give clues about the binary geopotential surface values. Our results also agree with \citet{Stern2019b}, who noted that the equatorial regions of the large and small lobes are binary geopotential highs.

\subsection{Surface Accelerations}
\label{sec:forc:acc}
Figure \ref{fig:forc_2} expresses the numerical acceleration computed over Arrokoth's surface in mm\,s$^{-2}$, where this arises from both the gravitational and centripetal accelerations and is computed using the gradient of Eq. \eqref{eq:grav_1}:
\begin{align}
|-\nabla V_b(x,y,z)|.
\label{eq:grav_9}
\end{align}
The acceleration has the highest values at the pole regions of the large and small lobes, and is much higher at the small lobe's poles (red). The surface acceleration has a local point of maxima value of $9.07 \times 10^{-1}$ mm\,s$^{-2}$ at the large lobe's poles, while it has a global point of maxima value of $9.50$ mm\,s$^{-2}$ at the small lobe's poles. The results imply that the large lobe has a slightly lower maximum acceleration value of $\sim 95.47\%$ than the small lobe. Therefore, the large and small lobes considerably influence the intensity of acceleration experienced by a dust particle near the surface of the body. For comparison purposes, the magnitudes of the accelerations computed on Pluto's surface\footnote{\url{https://nssdc.gsfc.nasa.gov/planetary/factsheet/plutofact.html}} are $620$\,mm\,s$^{-2}$, about six hundred times greater than the magnitudes of Fig. \ref{fig:forc_1}. For Earth's surface acceleration ($9,800$\,mm\,s$^{-2}$), we have an approximate factor ten thousand times greater. The Arrokoth numerical \textit{surface accelerations} range computed by our polyhedral model approximated between $0.55-0.95$ mm\,s$^{-2}$ (Fig. \ref{fig:forc_2}) from a global point of minima at the neck (blue) to a global point of maxima at the small lobe's poles (red), which is also in accordance with the estimated Arrokoth surface acceleration limits obtained by \citet{Stern2019b} and \citet{McKinnon2020}. Again, the centripetal potential influence is not significant. This can be confirmed when comparing the accelerations found in Fig. \ref{fig:grav_3} from $|-\nabla U_b|$. Figure \ref{fig:forc_2} is also in line with Fig. \ref{fig:forc_1}: i.e., the surface locations of minimum binary geopotential values (poles) correspond to sites of maximum surface acceleration values (poles). The regions on Arrokoth's surface with maximum binary geopotential values (equator) correspond to locations of minimum surface acceleration values (equator).

\subsection{Surface Slopes}
\label{sec:forc:slo}
The mapping of \textit{surface slope} angles assist in understanding the motion of free particles across the surface of the body. Equation \eqref{eq:grav_10} defines the surface slope angles in correlation with the local topography and local binary geopotential field:
\begin{align}
\theta = 180^\circ - \acos(\dfrac{-\nabla V_b(x,y,z)\cdot\hat{n}}{|-\nabla V_b(x,y,z)|}).
\label{eq:grav_10}
\end{align}
By definition, if slope angles $\theta>90^\circ$, then the local surface faces are unstable sites, where cohesionless particles will be ejected from the body. In contrast, for surface slope angles in a range of $0^\circ<\theta<90^\circ$, the movement of free particles is defined by the friction angle $\theta_f$, where $\mu_f=\tan(\theta_f)$ is the coefficient of friction.
\begin{figure}
  \centering
  \fbox{\includegraphics[width=8.44cm]{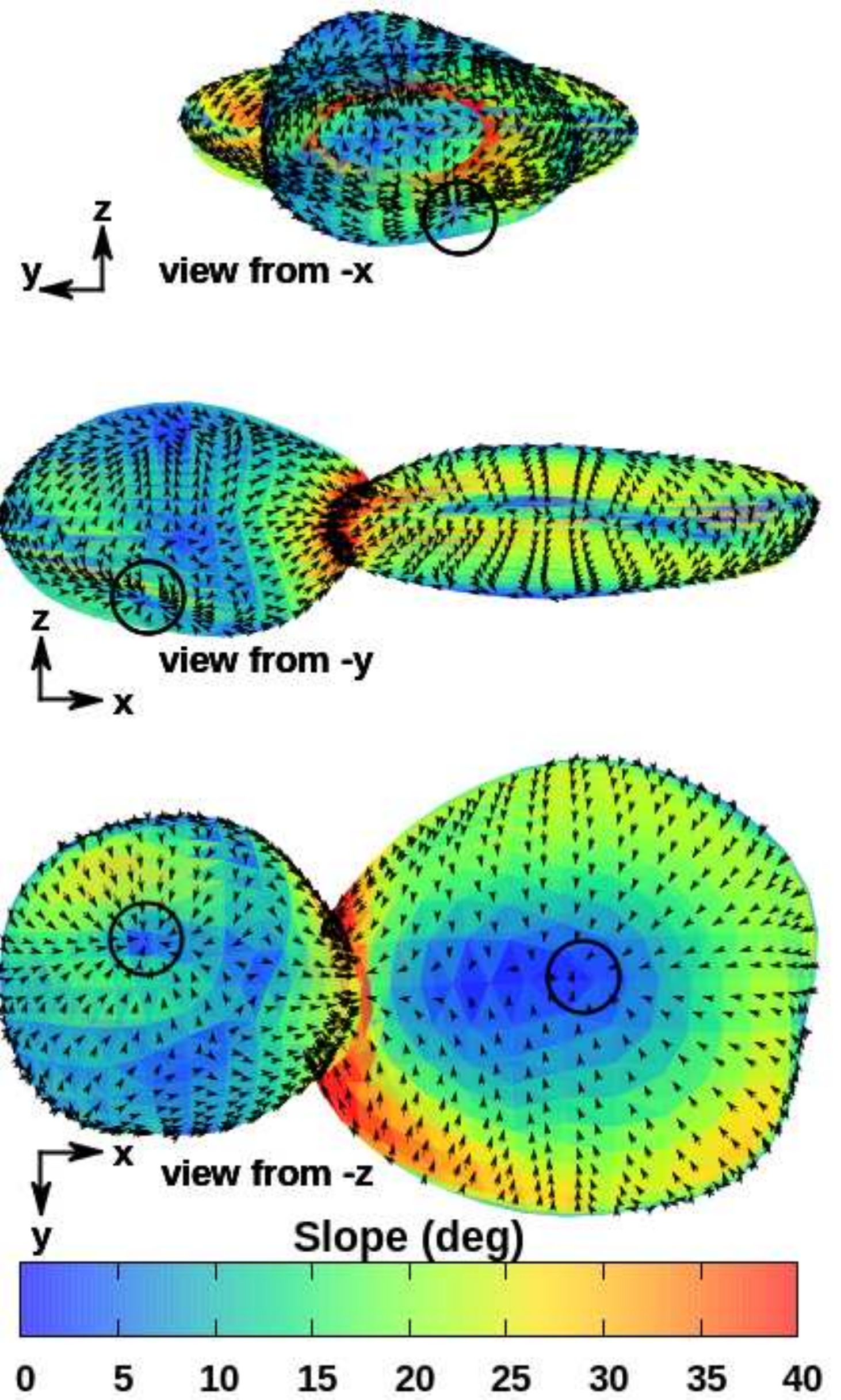}}
  \fbox{\includegraphics[width=8.44cm]{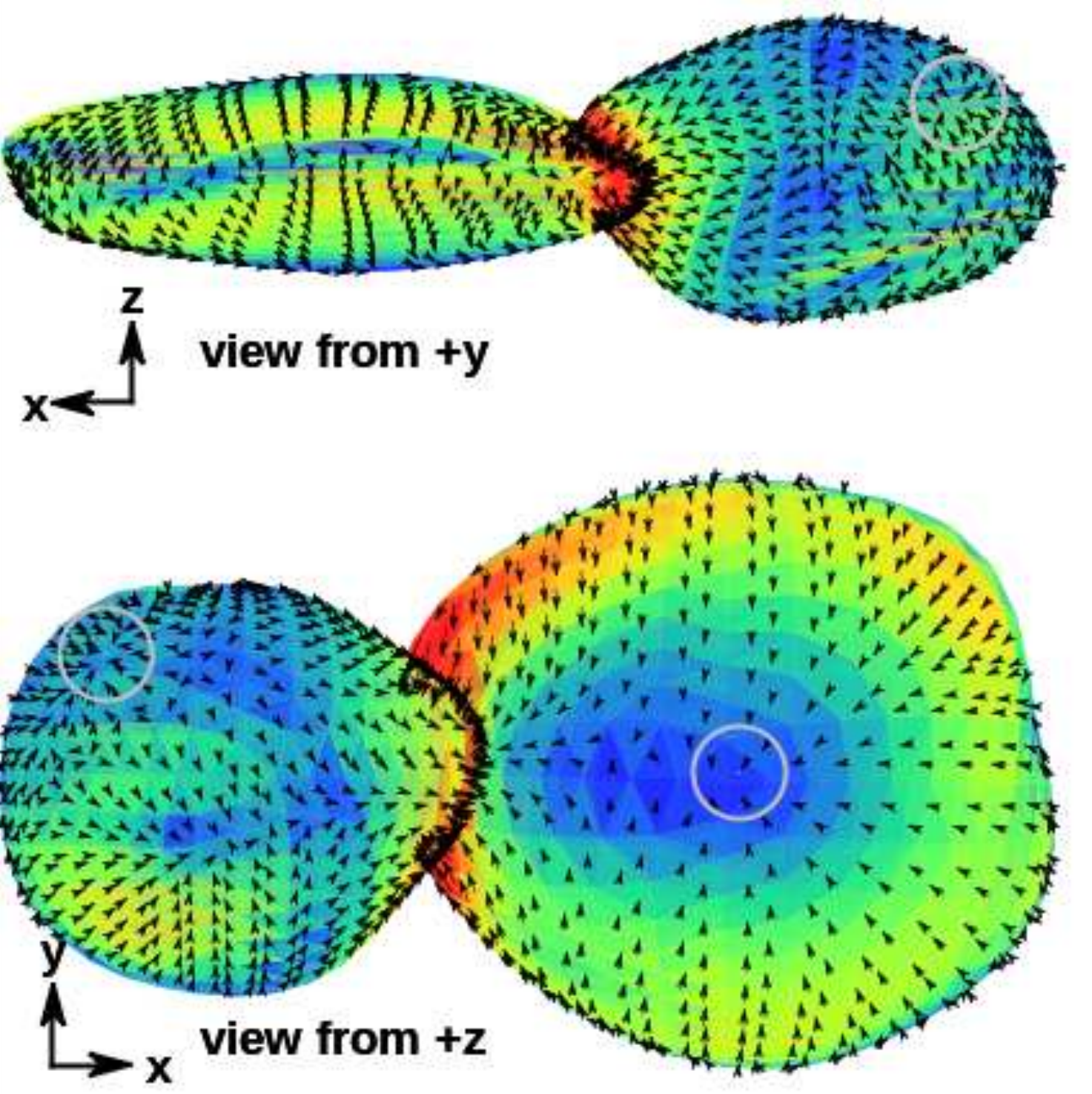}}
  \caption{Directions of the local acceleration vector field tangent to the surface of Arrokoth contact binary ($\rho=0.5$\,g\,cm$^{-3}$). The color code denotes surface slope angles, in degrees.}
  \label{fig:forc_4}
\end{figure}
\begin{figure}
  \centering
  \fbox{\includegraphics[width=8.44cm]{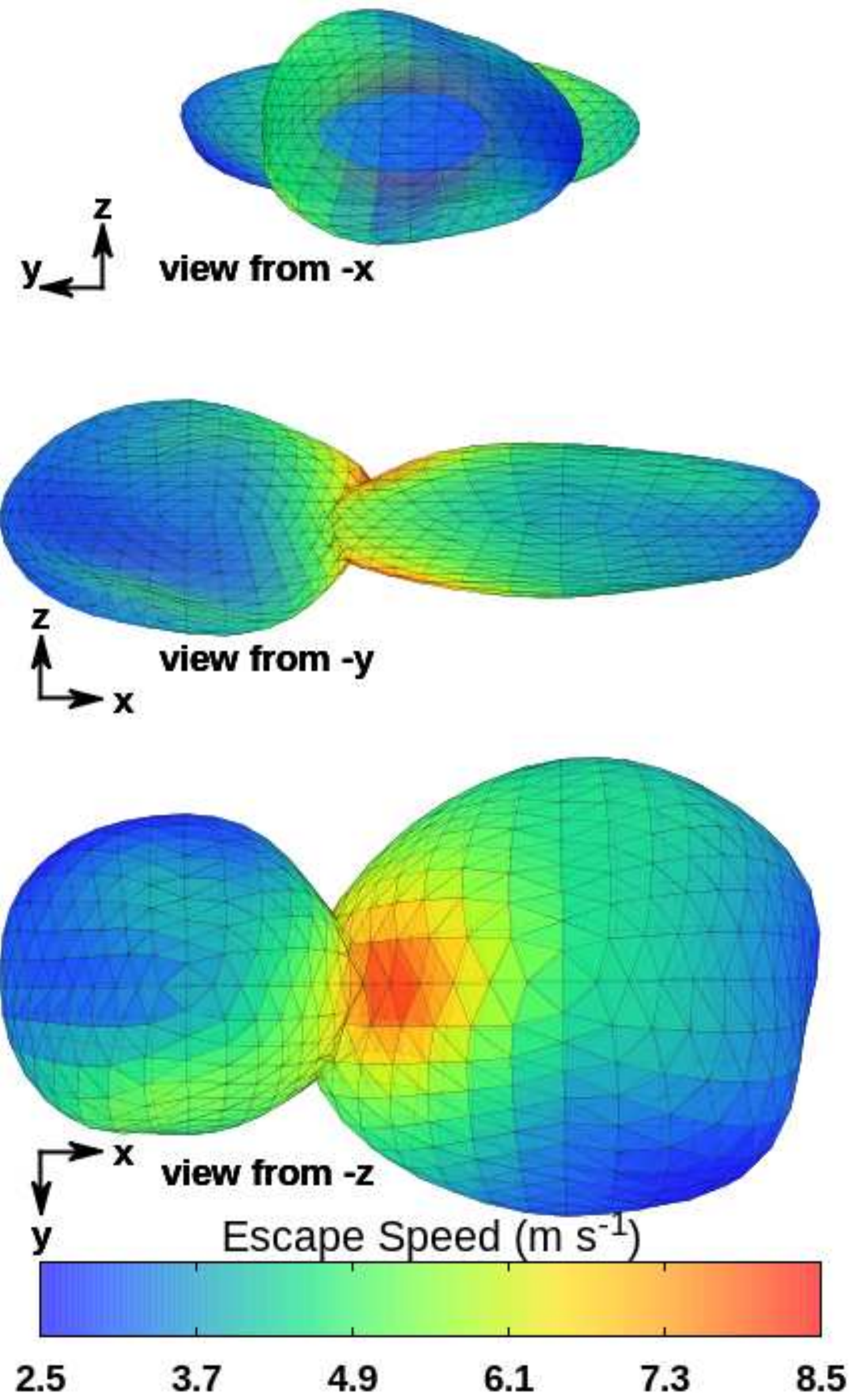}}
  \caption{Local normal escape speed over Arrokoth contact binary surface, in m\,s$^{-1}$. If the launch speed is greater than the given speed the particle will escape. These values are derived assuming the particle is launched normally to the surface.}
  \label{fig:forc_5}
\end{figure}
Figure \ref{fig:forc_4} shows the slope angles over the Arrokoth lobes' surfaces. The colour panel shows that the slope angles over the surface of Arrokoth contact binary are low, except in the neck, where they can exceed $40^\circ$ and reach a maximum slope angle of $146^\circ$ (which occurs at the longitude of $134^\circ$ and latitude of $2^\circ$). The minimum slope angle is close to zero and it lies over the large lobe poles $\sim 6$\,km apart from the barycentre. Given the friction angle $\theta_f=40^\circ$, a friction coefficient $\mu_f$ of at least $0.84$ would be necessary for a non-sliding condition to hold over the entire surface of each of Arrokoth's lobes. When the local surface slope angle $\theta$ exceeds the friction angle $\theta_f$, the particles that were initially at rest start moving towards the locations of the lower intensity of surface slopes. While the portion of the body surface corresponding to $\theta<\theta_f$ is a stable site that allows particles to attach and is susceptible accumulation of materials. From the perspectives $-y$ and $-z$ of Fig. \ref{fig:forc_4}, we note that most of the highest slope angle values are located across the surface of the large lobe (between $15^\circ$ and $40^\circ$), while most of the slope values over the small lobe's surface are small (between $0^\circ$ and $15^\circ$). Although the maximum value of the slope angle mapped on the surface of Arrokoth's lobes is located at the neck, the large lobe's equator also has a considerable number of sites close to high slope intensity, which is shown by the yellow to red colour box range in the equatorial area located at the beginning of the large lobe's equator near the neck region. Comparing the large and small lobes from Fig. \ref{fig:forc_4}, Arrokoth's dynamic slope angle $\theta$ is longitudinally uniform across the large lobe and increasing in the latitudinal directions from the poles to the equator, while the small lobe shows some dynamic peculiarities of slope angles that emerge from its longitudinal non-uniformity. In general, except at the neck, Figs \ref{fig:forc_1}, \ref{fig:forc_2} and \ref{fig:forc_4} show that the sites that comprise the lowest binary geopotential intensities also indicate the lowest dynamic slope angle values. Locations that have the highest binary geopotential values, also have the highest slope intensities. Therefore, the flow tendency of surface particles across the surface of each Arrokoth lobe is to migrate towards the locations that cover the lowest intensities of the dynamic slope angle.

\subsection{Surface Stability}
\label{sec:forc:ss}
To validate the analysis in the previous subsection, we considered the centripetal potential in illustrating the directions of the tangential gravity attraction vector field for each Arrokoth lobe. In general, the vectors point in the direction of downslope motion. Figure \ref{fig:forc_4} shows the tangential acceleration vector field for the large and small lobes. The colour box code represents the dynamic slope angle $\theta$. The arrows over the large lobe are clearly mostly pointing from the equatorial region towards the polar regions, and also towards the neck. In this case, the equatorial region is not suitable for containing loose material, while the polar locations have some sink areas that would retain surface particles. This is a peculiar slope pattern, which is due to Arrokoth contact binary's high spin period. For minor bodies, which in general have a sufficiently lower rotation period, the slope arrows are reversed and the equatorial region is the stable region. We verified this reversal of the expected behavioural pattern, e.g., in asteroid triple systems (2001) SN$_{263}$ Alpha \citep{Winter2020}, binary systems (1999) KW$_4$ Alpha \citep{Scheeres2012} and single systems (101955) Bennu \citep{Scheeres2016}. The central region (black and grey circles) of the large lobe's poles carry the lowest slope angle values at well below $5^\circ$. From perspective $\pm z$, there is a yellow area between the poles and the small lobe's equator that have a slightly higher slope intensity, which shows the existence of local surface depressions observed by \citet{Spencer2020}. The results are in agreement with the surface tilt angles of the small lobe in Fig. \ref{fig:prop_5}. The black circles across the small lobe's surface (perspectives $-x$, $-y$ and $-z$) indicate the same sink area for downslope particle motion direction. There is a stable resting site at this location that would be favourable to the accumulation of materials, which would explain the salience, i.e., a kind of bulge (left-hand side of Fig. \ref{fig:prop_3}) between the poles and the equatorial regions of the small lobe. This same behaviour can be evidence for another sink area on the small lobe represented by grey circles in perspectives $+y$ and $+z$.

Another peculiarity about the dynamic slope angle $\theta$ is the proximity of the neck. In this region, we have an upslope motion. This occurs because the surface tilt angles at the neck have the highest values (Fig. \ref{fig:prop_5}). Then, the supplementary angle between the normal face vector and the total gravity attraction vector is greater than $90^\circ$, i.e., $\theta>90^\circ$. Hence, the neck is an unstable region in which a free particle can be removed from Arrokoth contact binary.
Nevertheless, the overall picture of the slope angles and tangential acceleration vector fields do not vary significantly for densities\footnote{An animated movie is available online that shows how the overall picture of the dynamic slope angles change as a function of density (Movie 2).} up to 0.25\,g\,cm$^{-3}$.

\subsection{Escape Speed}
\label{sec:forc:espeed}
Another item of interest for surface stability is the necessary \textit{escape speed} that a loose particle experiences over Arrokoth's lobes, which defines the boundary numbers for the local launch speed below which the free particles may re-impact Arrokoth contact binary's surface. Let us compute the escape speed considering only the local gravity, Arrokoth's total mass and its rotational motion \citep{Scheeres2012}. Then, the escape speed $v_e$ yields:
\begin{align}
v_e = -\hat{n}\cdot(\pmb{\Omega}\times\textbf{r})+\sqrt{[\hat{n}\cdot(\pmb{\Omega}\times\textbf{r})]^2 -2U_{b_{min}}-(\pmb{\Omega}\times\textbf{r})^2}
\label{eq:grav_11}
\end{align}
\noindent where \textbf{r} is the radius vector from the body's centre mass to the local surface and $U_{b_{min}}=min\bigg[U_b,-\frac{GM}{|\textbf{r}|}\bigg]$.

From Eq. \eqref{eq:grav_11}, we compute the escape speed $v_e$ over Arrokoth's lobe surface. Figure \ref{fig:forc_5} shows the escape speed mapped across the surfaces of each of Arrokoth's lobes using the binary gravitational force potential $U_b$ from the polyhedral model of Arrokoth (Fig. \ref{fig:grav_2}). On the one hand, the launch speed numbers are distributed mostly with low values over Arrokoth's lobes. On the other hand, the high intensities (up to $8.9$\,m\,s$^{-1}$) of escape speeds are located over the hemisphere of the large lobe and near the neck. Over most of the surface, the escape speed is between $2.5-8.5$\,m\,s$^{-1}$, and in the neck between Arrokoth's two lobes, at its perspective $\pm z$, the escape speeds can achieve $7.9$\,m\,s$^{-1}$. It is important to note that escape speeds are not a well-defined quantity when Arrokoth contact binary has locally non-convex regions, where a speed normal for the surface would result in a re-impact with a different surface location, which is the case for the neck region. Nevertheless, the results still indicate the level of speed generally necessary to generate energies consistent with escape from Arrokoth contact binary.

\section{Equilibrium Points}
\label{sec:eq}
Arrokoth contact binary's \textit{equilibrium points} are the critical points of the binary geopotential $V_b(x,y,z)$. Thus, considering Arrokoth's binary geopotential (Eq. \eqref{eq:grav_1}), the location of these equilibrium points can be found by solving the following equation:
\begin{eqnarray}
   -\nabla  V_b(x,y,z) = \textbf{0}.
\label{eq:eq_1}
\end{eqnarray}
For each lobe, we can compute their equilibrium points taking the gradient of each single geopotential $V_l(x,y,z) = -\frac{1}{2}\omega^2(x^2+y^2)+U_l(x,y,z)$, where $l=1,2$. 
\begin{table*}
\centering
  \caption{Location of equilibrium points about Arrokoth, large and small lobes, and their values of latitude $\phi$, longitude $\lambda$, radial barycentre distance, and the geopotential, $V_m(x,y,z)$, $m=b,1,2$. Equilibria are computed through polyhedral or mascons techniques with Minor-Equilibria package and an accuracy of $10^{-5}$, assuming a constant density of $\rho=0.5$\,g\,cm$^{-3}$ and spin period. We use a rotation period of $15.92$\,h for Arrokoth and $9.2$\,h for each lobe.}
 \label{tab:eq_1}
 \scalebox{1.0}
{
 \begin{tabular}{cccccccc}
  \toprule
  Point & X (km) & Y (km) & Z (km) & $\phi$ (deg) & $\lambda$ (deg) & radii (km) & $V_m(x,y,z)$ (m$^2$\,s$^{-2}$) \\
  \hline
  \multicolumn{8}{c}{Arrokoth} \\
  $E_1$ & 22.4071 & -0.561209 & 0.00426405 & 0.0108999 & 358.565 & 22.4142 & -7.31413 \\
  $E_2$ & -1.61385 & 17.2582 & 0.0371507 & 0.122802 & 95.3423 & 17.3335 & -6.08598 \\
  $E_3$ & -22.7559 & 0.0167606 & 0.0155456 & 0.0391414 & 179.958 & 22.7559 & -7.39587 \\
  $E_4$ & -1.36571 & -17.2629 & -0.00435036 & -0.0143939 & 265.477 & 17.3169 & -6.08402 \\
  $E_5$ & -1.36938 & -0.0329298 & -0.0427635 & -1.78816 & 181.377 & 1.37044 & -10.9875 \\
  $E_6$ & -8.26302 & 0.0349151 & -0.0119711 & -0.0830067 & 179.758 & 8.26310 & -11.6038 \\
  $E_7$ & 5.75681 & 0.00501069 & 0.0211256 & 0.210256 & 0.0498698 & 5.75685 & -11.4559 \\
  \hline
  \multicolumn{8}{c}{Large lobe} \\
  $L_1$ & 0.339640 & 10.6940 & 0.287504 & 1.53922 & 88.1809 & 10.7033 & -5.88166 \\
  $L_2$ & 9.51317 & -7.81606 & -0.113458 & -0.527967 & 320.593 & 12.3128 & -6.53292 \\
  $L_3$ & 2.43243 & -10.8101 & 0.168365 & 0.870533 & 282.681 & 11.0817 & -6.25730 \\
  $L_4$ & -6.29143 & -10.6187 & -0.162600 & -0.754767 & 239.354 & 12.3436 & -6.62660 \\
  $L_5$ & 0.0294675 & 0.0356299 & 0.0845775 & 61.3356 & 50.4078 & 0.0963907 & -8.18868 \\
  \hline
  \multicolumn{8}{c}{Small lobe} \\
  $S_1$ & -1.84050 & -9.06460 & -0.356543 & -2.20749 & 258.522 & 9.25643 & -4.56647 \\
  $S_2$ &  5.63226 &  8.09837 & 0.0994874 & 0.577838 & 55.1821 & 9.86488 & -4.88852 \\
  $S_3$ & -3.55255 &  8.44461 & -0.502648 & -3.14041 & 112.816 & 9.17523 & -4.69751 \\
  $S_4$ & -9.16026 &  3.70767 & -0.182808 & -1.05978 & 157.964 & 9.88386 & -4.82960 \\
  $S_5$ & -0.0941824 & -0.453653 & -0.300980 & -33.0080 & 258.271 & 0.552503 & -6.92815 \\
  \hline
 \end{tabular}}
 \end{table*}
\begin{figure}
  \centering
  \fbox{\includegraphics[width=8.44cm]{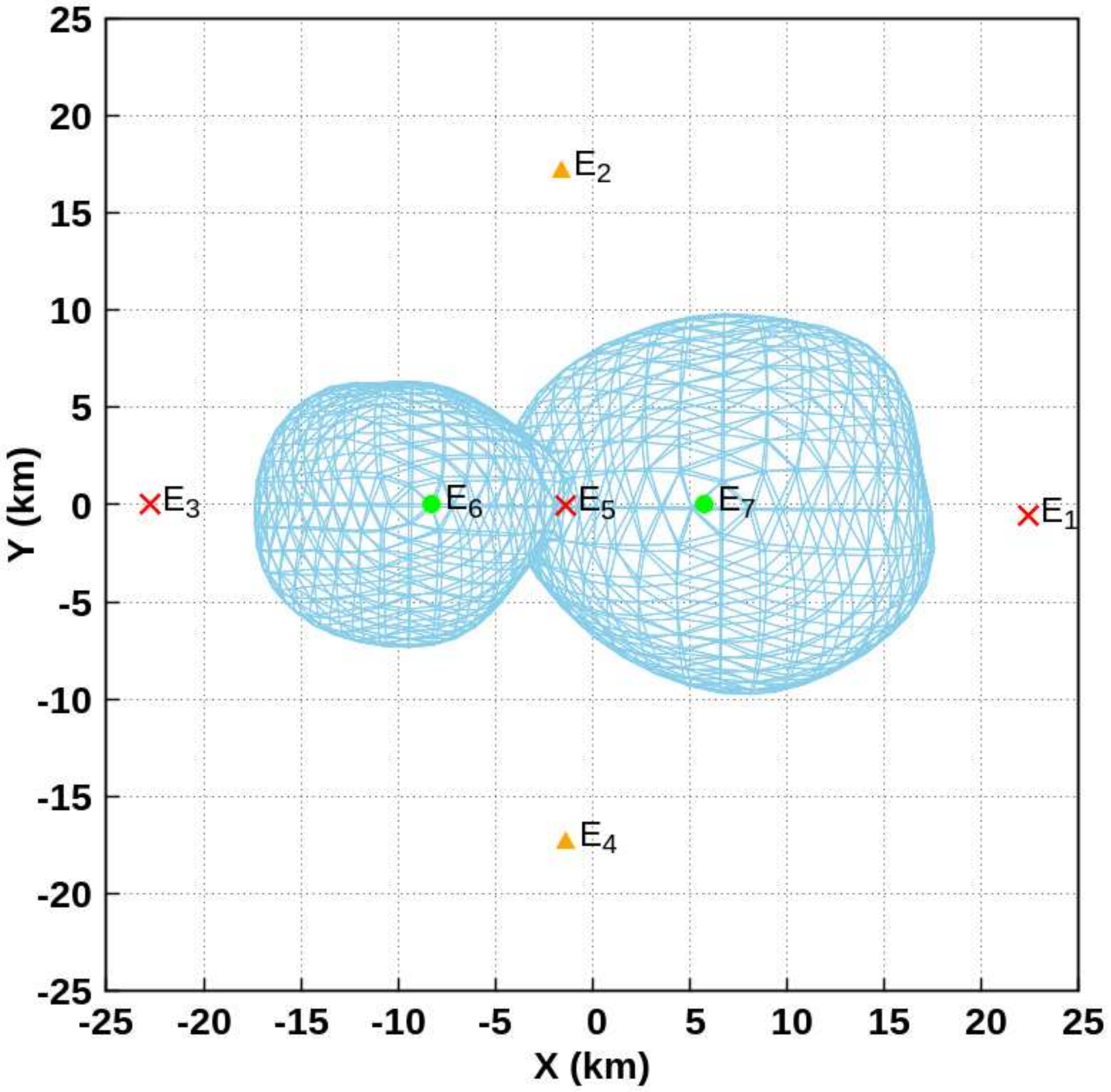}}
  \caption{Location of the seven equilibrium points of the Arrokoth contact binary in the projection plane $xOy$ and for a density of $\rho=0.5$\,g\,cm$^{-3}$. Red X-dots are topologically classified as saddle--centre--centre points (hyperbolically unstable), orange triangular-dots as sink-source-centre points (complexly unstable) and green circle-dots as centre--centre--centre points (linearly stable). (For
interpretation of the references to color in this figure legend, the reader is referred to the web version of this article.)}
  \label{fig:eq_1}
\end{figure}
\begin{figure*}
  \centering
  \fbox{\includegraphics[width=8.44cm]{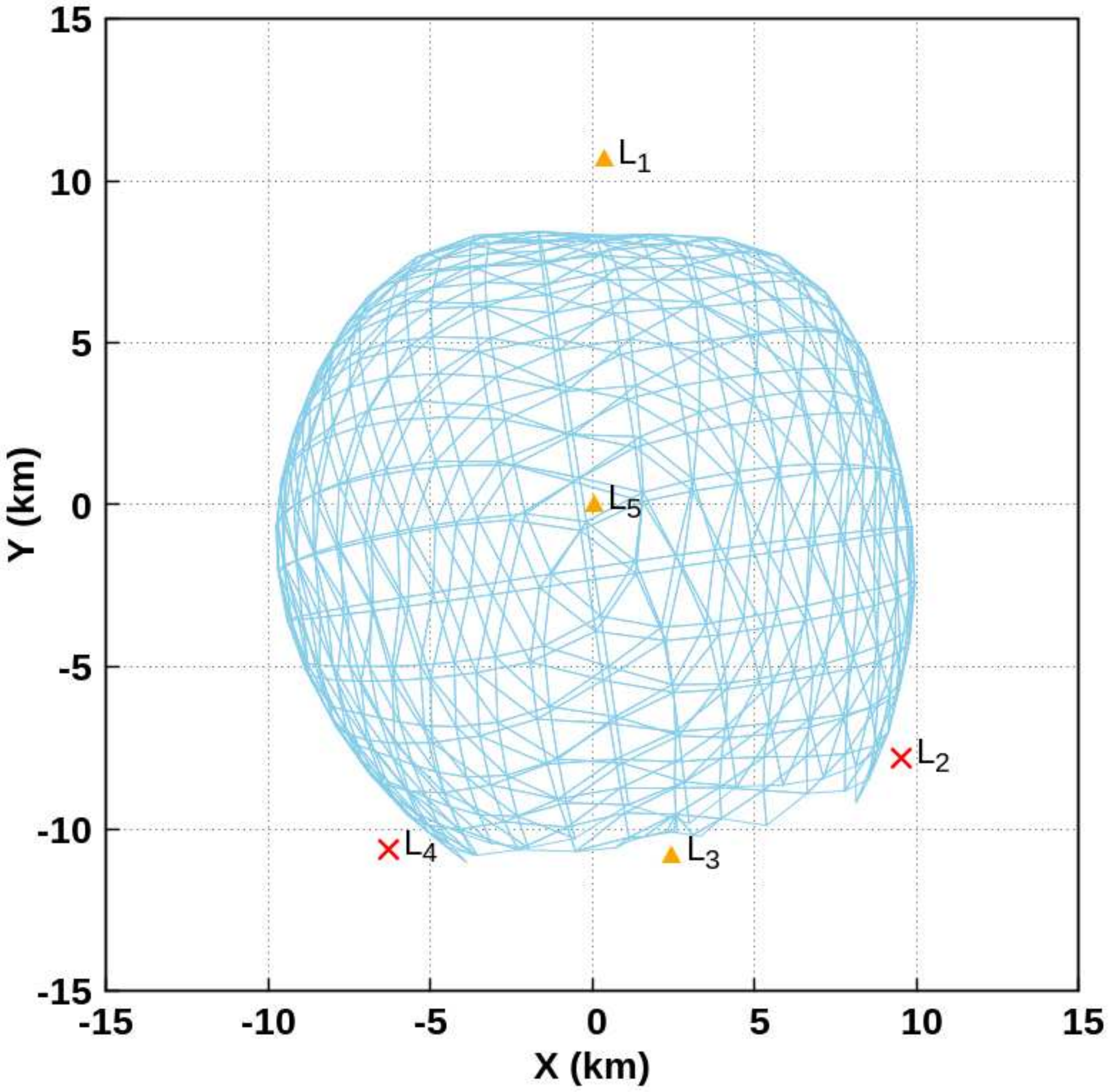}}
  \fbox{\includegraphics[width=8.44cm]{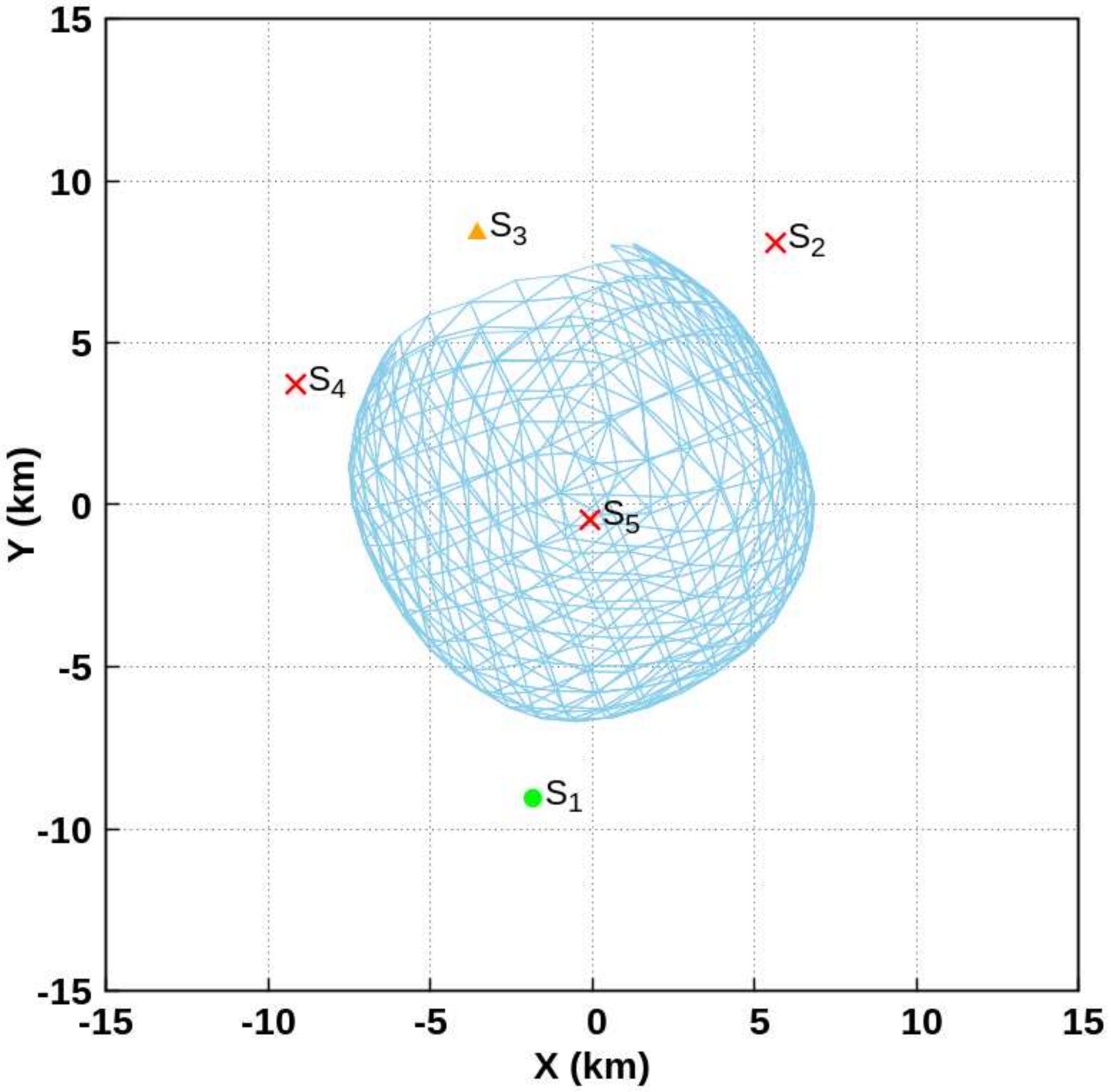}}
  \caption{Location of the five equilibrium points of the large (left-hand side) and the small (right-hand side) lobes seen from their $xOy$ projection planes, for a density of $\rho=0.5$\,g\,cm$^{-3}$ and a rotation period of $9.2$\,h.}
  \label{fig:eq_2}
\end{figure*}
\subsection{Equilibria Location}
\label{sec:eq:eqloca}
Our computed equilibrium points for Arrokoth and its individual lobes are presented in Table \ref{tab:eq_1}. The equilibria stability can be investigated using quantities such as \textit{zero-velocity curves}, \textit{return speed}, \textit{orbital energy}, and the \textit{gravity-power} equation. The equilibrium points are stationary orbits in the binary body-fixed co-ordinate frame and the number of solutions of Eq. \eqref{eq:eq_1} depend on the shape and spin period of the body. Our polyhedral model derived from \citet{Stern2019b} provided seven equilibrium points for Arrokoth's shape with a density of $\rho=0.5$\,g\,cm$^{-3}$, $1,952$ triangular faces and a spin rate of 15.92\,h. Using the polyhedral technique, we were found four external equilibrium points ($E_1$, $E_2$, $E_3$ and $E_4$) and three inner equilibrium points ($E_5$, $E_6$ and $E_7$). Table \ref{tab:eq_1} shows the location of Arrokoth's equilibrium points. Additionally, we indicate their latitude $\phi$ (degrees), longitude $\lambda$ (degrees), radial distance from centre mass (km) and geopotential $V_m(x,y,z)$, $m=b,1,2$ (m$^{2}$\,s$^{-2}$) values. Table \ref{tab:eq_1} shows that all of Arrokoth's seven equilibrium points are slightly out-of-plane. In other words, they are not in the equatorial plane $xOy$ because of its asymmetrical shape in the latitudinal direction (left-hand side of Fig. \ref{fig:prop_3}). The equilibrium points $E_1$ and $E_4$ have the same order of magnitude in their $z$-axis components ($\sim 10^{-3}$), while the other points are one order of magnitude higher ($\sim 10^{-2}$). We also note that equilibrium points $E_5$ and $E_7$ are located inside the large lobe, while $E_6$ is inside the small lobe. Equilibrium point $E_5$ is the closest to Arrokoth's centroid at approximately $1.37$\,km away. This behaviour can be more efficiently observed if we examine Fig. \ref{fig:eq_1}, where we plotted the arrangement of all of Arrokoth's equilibrium points in its projection plane $xOy$. This figure illustrates another peculiarity of Arrokoth's equilibrium points. Due to Arrokoth's high ellipticity, there is no radial symmetry on its external equilibrium points, only axial symmetry. As can be seen from the radii column of Table \ref{tab:eq_1}, axial symmetry occurs with the pairs of equilibrium points $E_1-E_3$ and $E_2-E_4$. Equilibrium point $E_2$ has an approximately $5^\circ$ left-hand side offset from the $y$-axis ($\lambda=95^\circ$), while equilibrium point $E_4$ also has a $5^\circ$ left-hand side offset from the $y$-axis ($\lambda=265^\circ$), i.e., both are at the same side of the $y$-axis and they are almost on the same line as $E_2-E_5-E_4$ (see X column), while equilibrium points $E_1$ and $E_3$ are at opposite sides of the $x$-axis ($\lambda=358^\circ,179^\circ$; respectively).

In addition, we divided Arrokoth contact binary's original polyhedral shape into two lobes using the neck's dimensions. We used 849 triangular faces with 4,463 mascons \citep{Geissler1996,Werner1997b,Scheeres1998} for the large lobe and 809 triangular faces with 3,127 mascons for the small lobe. We then used the Minor-Equilibria package with the mascons approach to find the location of the equilibrium points for each lobe. Each lobe has its own body-fixed co-ordinate frame for describing the locations of equilibrium points. We also assume that each individual lobe has a uniform rotation about its own largest moment of inertia ($z$-axis) along with their centroids with same spin period of 9.2\,h, used for the rubble-pile model for Arrokoth's solar nebula origin \citep{McKinnon2020} and near the rotation period of 10\,h used by \citet{Grishin2020}, for smoothed particles hydrodynamic simulations of the origin of Arrokoth-like Kuiper-belt contact binaries from wide binaries. All equilibrium points are computed for a constant density of $\rho=0.5$\,g\,cm$^{-3}$. Our results of the equilibria from the large and small lobes are shown in Table \ref{tab:eq_1} and plotted in Fig. \ref{fig:eq_2}. In Table \ref{tab:eq_1}, note that we use the same co-ordinates notation X, Y and Z to represent the locations of equilibria for Arrokoth and its large and small lobes, although they are on different frames. We found five equilibrium points for each individual lobe, considering their shape and spin rate. We denoted the equilibrium points of the large and small lobes by $L_i$ and $S_i$ ($i=1,...,5$), respectively. Note from Fig. \ref{fig:eq_2} that if the large and small lobes are merged, then the location of equilibrium point $L_2$ is relatively close to the location of equilibrium point $S_2$, which suggests that they are created from Arrokoth equilibrium point $E_2$ (Fig. \ref{fig:eq_1}). This same behaviour can also be observed from the pairs of equilibrium points $L_3$/$S_3$ and $L_4$/$S_4$ that are relatively close to each other; thus, they would be created from Arrokoth equilibrium points $E_5$ and $E_4$, respectively (see, e.g., \citet{Jiang2015,Yu2018}). We used a different approach to these authors. We computed the gravitational field of each individual lobe separately and the lobes do not interact with each other. In addition, from Table \ref{tab:eq_1}, we can see that equilibrium points of the large and small lobes preserve the radial asymmetry of the original Arrokoth shape due to their elongation.

\subsection{Equilibria Stability}
\label{sec:eq:eqsta}
We also examined the stability of the equilibria shown in Table \ref{tab:eq_1}. We studied the linear stability of the equilibrium points. The unnormalized eigenvalues for each given equilibrium point and their corresponding topological stabilities are shown in Table \ref{tab:eq_2} (Appendix \ref{sec:eigen}). According to the eigenvalues, the first five equilibrium points, i.e., $E_1$, $E_2$, $E_3$, $E_4$ and $E_5$, are unstable. However, they have a different topological stability. The odd indices of the equilibrium points have a saddle--centre--centre topological structure (hyperbolically unstable), while the even indices are associated with a sink--source--centre stability (complexly unstable). We show that equilibrium point $E_5$ is unstable and near Arrokoth's centre mass. In addition, equilibrium points $E_6$ and $E_7$ have a centre--centre--centre topological structure that is linearly stable. Following \citet{Scheeres1994}'s definition of external equilibrium points, Arrokoth contact binary can be classified as a minor body of type II. The large lobe alternates between sink--source--centre and saddle--centre--centre topological structures in a different type of Arrokoth contact binary's equilibria. The odd indices are complexly unstable points ($L_1$, $L_3$ and $L_5$), while even indices ($L_2$ and $L_4$) are hyperbolically unstable points. As a sink--source--centre point, the central equilibrium point $L_5$ also has a different topological structure to Arrokoth's central point. The small lobe does not have a sequential pattern. It has one unstable point identified as sink--source--centre ($S_3$), three equilibrium points with saddle--centre--centre topological stability, including the central point ($S_2$, $S_4$ and $S_5$) and a single external linearly stable point ($S_1$).

\citet{McKinnon2020} inferred that the large and small lobes were already aligned before the final merge and perhaps were in a mutually circular synchronous orbit. Thus, we additionally show the equilibrium points for each individual lobe in the rotation period of $15.92$\,h in Fig. \ref{fig:eq_3b}. For the adopted rotation period, the number of equilibria of the large and small lobes and also the topological structure of the large lobe did not change. However, the topological stability of external equilibrium point $S_3$ of the small lobe differs from that shown in Table \ref{tab:eq_2}, with a spin period of $9.2$\,h. For this rotation period, the external equilibrium point $S_3$ becomes linearly stable. In addition, for the rotation period of $9.2$\,h the large and small lobes both have a type II signature. Nevertheless, for the spin period of $15.92$\,h the large lobe maintains its type II signature, while the small lobe changes to a type I signature \citep{Scheeres1994}.

The geopotential intensity can also help to understand the stability of a determined equilibrium point. For example, Arrokoth inner equilibrium point $E_6$ has the lowest binary geopotential intensity (Table \ref{tab:eq_1}). Using the value of the binary geopotential to list the equilibrium points from small to large, we get $E_6$, $E_7$, $E_5$, $E_3$, $E_1$, $E_2$, as well as $E_4$. Thus, one can conclude that $E_4$ is the most unstable equilibrium point and $E_3$ is the least unstable external equilibrium point. Meanwhile, for the large lobe we can conclude that equilibrium point $E_1$ is the most unstable.

\subsection{Location and Stability of Equilibria at Different Density Values}
\label{sec:eq:dens}
\begin{figure}
  \centering
  \fbox{\includegraphics[width=8.44cm]{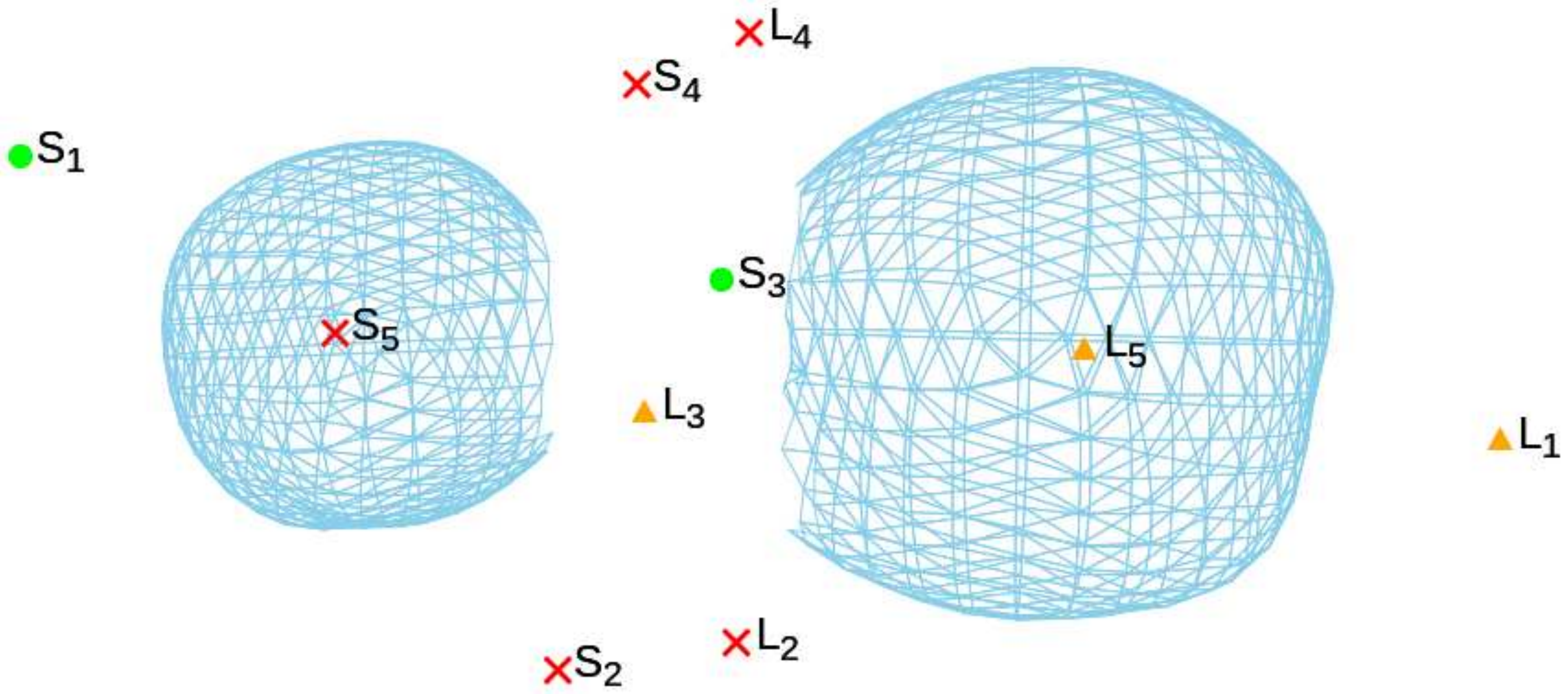}}
  \caption{Location of the five equilibrium points of the large ($L_i$) and small ($S_i$) lobes ($i=1,...,5$). We use a density of $\rho=0.5$\,g\,cm$^{-3}$ and a rotation period of $15.92$\,h. Red X-dots are topologically classified as saddle-centre-centre points (hyperbolically unstable), orange triangular-dots as sink--source--centre points (complexly unstable), and green circle-dots as centre--centre--centre points (linearly stable).}
  \label{fig:eq_3b}
\end{figure}
\begin{figure}
  \centering
  \fbox{\includegraphics[width=8.44cm]{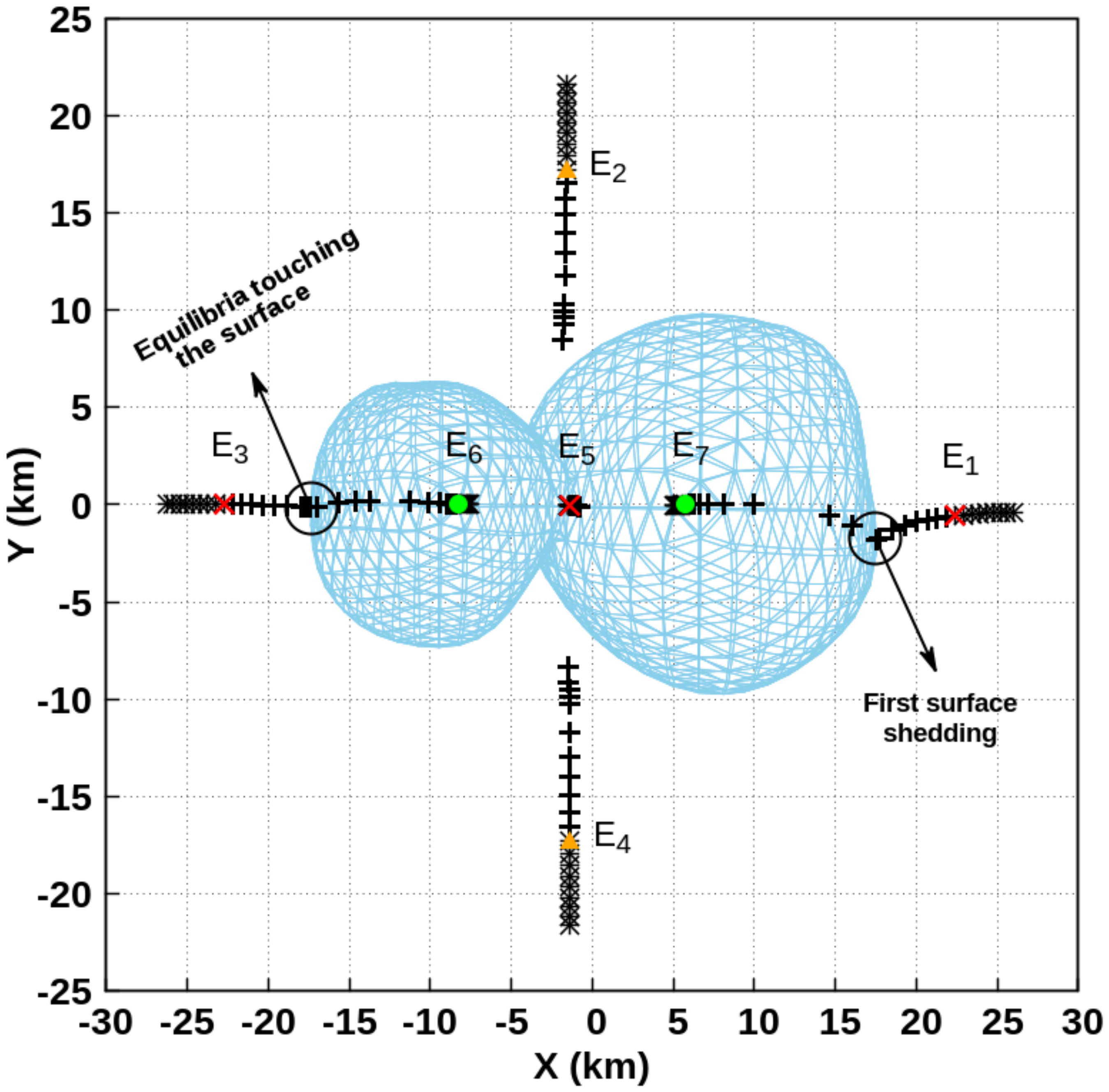}}
  \caption{Equilibrium points of the contact binary Arrokoth in the projection plane $xOy$ for a set of density values. Black marks $\pmb{+}$ and $\pmb{\ast}$ show the equilibrium points trajectories and they are computed from down and up density values, respectively, in the range of $0.5\pm 0.4$\,g\,cm$^{-3}$.}
  \label{fig:eq_1b}
\end{figure}
Arrokoth contact binary's density and mass are based on assumptions \citep{Stern2019b}; therefore, we analyzed the effect of different densities on the equilibrias dynamic features. In addition, the equilibria can be changed in terms of their number, location and stability. For example, \citet{Feng2016} applied a dynamic model of a contact binary body consisting of two lobes (ellipsoidal and spherical) that are in physical contact in the contact binary system (1996) HW1. They showed that equilibrium point $E_3$ can transit from a linearly stable topology to an unstable topology, with the decrease of a dimensionless scaling parameter that represents the ratio of the gravitational acceleration to centripetal acceleration. In our polyhedral model, we choose a range of $0.5\pm 0.4$\,g\,cm$^{-3}$ for $\rho$ \citep{McKinnon2020}, keeping the same volume and spin rate because these Arrokoth contact binary characteristics are more precisely known than its density \citep{Stern2019b}. We gradually varied Arrokoth's density using the upper and lower density values $0.5$\,g\,cm$^{-3}$, as shown in Figure \ref{fig:eq_1b} by the dynamic equilibria trajectories. The coloured equilibrium points divide these dynamic trajectories into two sets: equilibrium points, represented by the black marks $\pmb{+}$, that are computed for lower density values; and equilibrium points that are computed for the upper density values, represented by black marks $\pmb{\ast}$. For the lower density values, the external equilibria draw closer to the body, while for higher density values, the external equilibrium points move far away from the body.

This observation can be explained by the inertial frame. As an example, we look into equilibrium point $E_2$, where there is a balance between the gravitational and centripetal accelerations. If Arrokoth's density is decreased, while keeping the same volume and spin rate, then Arrokoth's gravitational acceleration at point $E_2$ is weak. Moreover, the gravitational acceleration is less than the centripetal acceleration at this point, and $E_2$ is no longer an equilibrium point. For a point far away from Arrokoth, the situation is even worse because the gravitational acceleration decreases with distance ($\propto 1/r^2$) and the centripetal acceleration will increase ($\propto r$). Thus, distant points show an absence of equilibria. To obtain a balance between the gravitational and centripetal accelerations in this case, we must consider dynamic equilibrium points close to Arrokoth. As the distance to Arrokoth becomes smaller, the gravitational acceleration will gradually increase in the dynamic trajectory for points towards Arrokoth. At a certain density, the balance between these two accelerations will be re-established to configure an equilibrium point. The same analysis can be performed for higher density values. In this case, the situation is reversed and the external equilibrium points move far away from Arrokoth ($\pmb{\ast}$), as shown in Fig. \ref{fig:eq_1b} and Movie 3. In addition, if Arrokoth's spin rate is changed instead of its density, the external equilibrium points will approach Arrokoth's surface for a faster spin rate. Meanwhile, at a slower spin rate, the external equilibrium points will move far away from Arrokoth.
\begin{figure}
  \centering
  \fbox{\includegraphics[width=8.44cm]{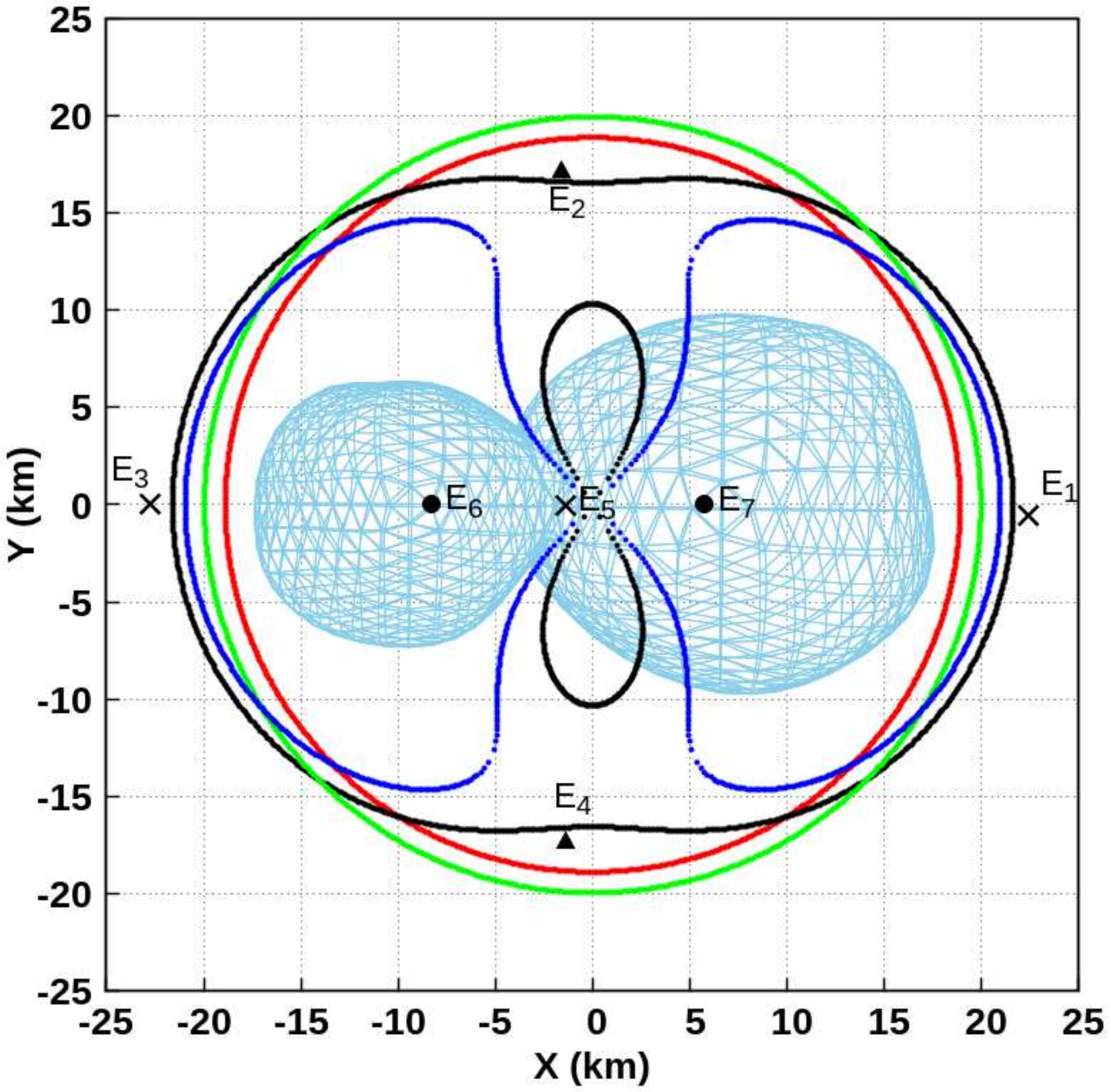}}
  \caption{1:1 Resonance radius $R_r$ shown in $xOy$ plane and computed from equation \eqref{eq:eq_6}. Color lines indicate a set of numerical combinations using two main $J_2$ and $C_{22}$ second-order and degree spherical harmonic coefficients from gravitational and centripetal accelerations. The color combinations are: no $J_2$ and no $C_{22}$ (red), $J_2+C_{22}$ (black), $J_2$ and no $C_{22}$ (green), and finally, $C_{22}$ and no $J_2$ (blue). The location of equilibrium points of the Arrokoth contact binary is represented by X-cross black dots. (For interpretation of the references to color in this figure legend, the reader is referred to the web version of this article.)}
  \label{fig:eq_4}
\end{figure}

However, the situation is different for the dynamic inner equilibrium points $E_6$ and $E_7$, which become close to the inner equilibrium point $E_5$ when the density increases and they follow their dynamic trajectories far away from it when the density decreases. In the case of lower density values, equilibrium points $E_6$ and $E_7$ move towards equilibrium points $E_3$ and $E_1$, respectively. When the densities are in the range of $0.13--0.14$\,g\,cm$^{-3}$, the equilibrium points $E_1$ and $E_7$ approach each other. At a certain density of this interval, $E_1$ and $E_7$ collide and annihilate each other ($E_1$/$E_7$) on Arrokoth contact binary's surface. After first annihilation, five equilibrium points remain ($E_2$, $E_3$, $E_4$, $E_5$ and $E_6$). At this point, Arrokoth's body should be below structural failure because this is the first surface-shedding condition for loose material to fly off its surface \citep{Hirabayashi2014}. As the density continues to decrease, $E_3$ and $E_6$ approach each other and touch Arrokoth's surface simultaneously. These two equilibrium points touch the same point on Arrokoth's surface and annihilate each other on the surface. After the second annihilation, for densities $\leq 0.11$\,g\,cm$^{-3}$, three equilibrium points remain ($E_2$, $E_4$ and $E_5$). Only $E_5$ is inside Arrokoth. Following \citet{Jiang2015,Yu2018}'s topological classification of Kleopatra-shaped objects, Arrokoth contact binary can be classified as topological case type I. The difference is that the first asteroid Kleopatra surface shedding occurs between equilibria pairs $E_3$/$E_6$ \citep{Hirabayashi2014}. Moreover, the first $E_1$/$E_7$ surface shedding appears asymmetrically over the large lobe's surface (below $x$-axis), because of its very irregular shape. 

The topological stabilities of equilibrium points do not change before the first and second annihilations (Table \ref{tab:eq_2}). However, depending on the range of densities and the spin rate adopted for a body, the equilibrium point stabilities can change (e.g., see \citet{Feng2016}). In comparing Figs \ref{fig:eq_1} and \ref{fig:eq_3b}, the equilibrium points of the large and small lobes did not preserve any topological stabilities from Arrokoth's original body (note that these figures are presented in the opposite perspective of views $\pm z$). The result suggests, e.g., that before the lobes were merged, i.e., before the equilibrium points $L_3$ and $S_3$ annihilated each other, the centre equilibrium point $E_5$ had two different topological structures: an unstable sink--source--centre from equilibrium point $L_3$ and a linearly stable one from equilibrium point $S_3$. Then, after merging, the central equilibrium point $E_5$ becomes an unstable saddle--centre--centre point. This same characteristic of topological stability can be noted for the other equilibrium points, e.g., the unstable equilibrium point $E_3$ can arise from the linearly stable $S_1$. This feature can also be observed even for the equilibrium points of the large and small lobes, as computed for the rotation period of 9.2\,h (Fig. \ref{fig:eq_2}).

Finally, our polyhedral model of Arrokoth shows a different number of equilibrium points from those found through ellipsoidal and spherical contact models \citet{Feng2016}. We found three more equilibrium points, i.e., the inner $E_5$, $E_6$ and $E_7$. Additionally, for the adopted range of densities, the equilibrium points $E_1$ and $E_3$ show the same topological stabilities (hyperbolically unstable) to those found by \citet{Feng2016}.

\subsection{Arrokoth Stability Through 1:1 Resonance}
\label{sec:eq:rb}
\citet{Hu2004} investigated the stability of orbital motion about a uniformly rotating arbitrary second-order and degree gravitational field. They stated the conservative bounds on zonal $J_2$ and tesseral $C_{22}$ gravity coefficients using the 1:1 resonance radius. They concluded that materials will be flung from an asteroid's surface when the 1:1 resonance radius intercepts it. We adapted their approach to consider the second-order and degree terms of the gravitational potential in the computation of the 1:1 resonance radius. Thus, we studied the influence of the $J_2$ and $C_{22}$ terms in Arrokoth's gravitational field for a density of $0.5$\,g\,cm$^{-3}$ and a spin period of $15.92$\,h. First, we need to find the 1:1 resonance radius, the radius where the centripetal acceleration from the body spin rate equals the point mass gravitational attraction of the body. We can find Arrokoth's 1:1 resonance radius using the following equation:
\begin{align}
r & = \dfrac{|-\nabla U(r,\lambda)|}{\omega^2},
\label{eq:eq_4}
\end{align}
\noindent where $|-\nabla U(r,\lambda)|$ represents the body acceleration truncated at second-order and degree, expressed as:
\begin{align}
|-\nabla U(r,\lambda)| & =\dfrac{GM}{r^2}\bigg[1+\dfrac{3}{2}J_2\bigg(\dfrac{R_s}{r}\bigg)^2 +9C_{22}\bigg(\dfrac{R_s}{r}\bigg)^2\cos(2\lambda)\bigg],
\label{eq:eq_5}
\end{align}
\noindent with point mass longitude represented by $\lambda$ and $R_s$ is the normalization radius defined previously.

If we suitably choose zonal $J_2$ and tesseral $C_{22}$ gravity coefficients so that they are normalized by squared 1:1 resonance radius $R_r$, then combining equations \eqref{eq:eq_4}-\eqref{eq:eq_5}, we can compute Arrokoth contact binary's 1:1 resonance radius $R_r$ as follows:
\begin{align}
R_r & = \sqrt[3]{\dfrac{GM}{\omega^2}\bigg[1+\dfrac{3}{2}\dfrac{J_2}{R_r^2} +9\dfrac{C_{22}}{R_r^2}\cos(2\lambda)\bigg]},
\label{eq:eq_6}
\end{align}
\noindent where we set $r=R_r$.

Figure \ref{fig:eq_4} indicates the influence of the zonal $J_2$ and tesseral $C_{22}$ gravity coefficients on Arrokoth's equilibrium points. The coloured lines show the 1:1 resonance radius $R_r$ and the different numerical experiments with these two coefficients. This figure shows the absence of second-order and degree gravitational perturbation terms, the 1:1 resonance radius (red line) is greater than the semi-axis $x$ with a mean radius value of $18.89$\,km. This fact occurs in most of the minor bodies, which have slow enough spin, so that $x<R_r$. However, if we include the combination of the second-order and degree terms of the gravitational acceleration (Eq. \eqref{eq:eq_5}), then the 1:1 resonance radius curves (black lines) intercept Arrokoth contact binary in some regions, i.e., $x>R_r$. This suggests that, at some sites, the body will be in tension and materials will be thrown off its surface (e.g., see Movie 2), as can be seen in the previous asteroidal analysis \citep{Hu2004}. Additionally, please note that the external black line follows the asymmetry of the external equilibrium points and it is very close to them. The black line also follows the equilibria by always keeping the equilibrium point in the outer region. Thus, we can infer that at this distance and beyond, the second-order and degree gravitational potential can be used as a good approximation for Arrokoth's real gravitational field. Figure \ref{fig:eq_4} shows that the $J_2$ 1:1 resonance radius (green line) does not intercept Arrokoth contact binary and the sink--source--centre points $E_2$ and $E_4$ are in the inner region of the 1:1 resonance radius, while the saddle--centre--centre points $E_1$ and $E_3$ are localized in the outer region. This shows that there is no radial symmetry about Arrokoth's external equilibrium points, only axial symmetry with a mean 1:1 resonance radius distance of $19.96$\,km. The 1:1 resonance radius $C_{22}$ (blue line) is most likely to appear close to unstable equilibrium points $E_1$, $E_3$, $E_5$, $E_6$, and $E_7$. We can see from Fig. \ref{fig:eq_4} that the inner equilibrium points are inside the blue line and the external equilibrium points are outside it. 

\subsection{Zero-Velocity Curves}
\label{sec:eq:zvs}
\begin{figure}
  \centering
  \includegraphics[width=8.6cm]{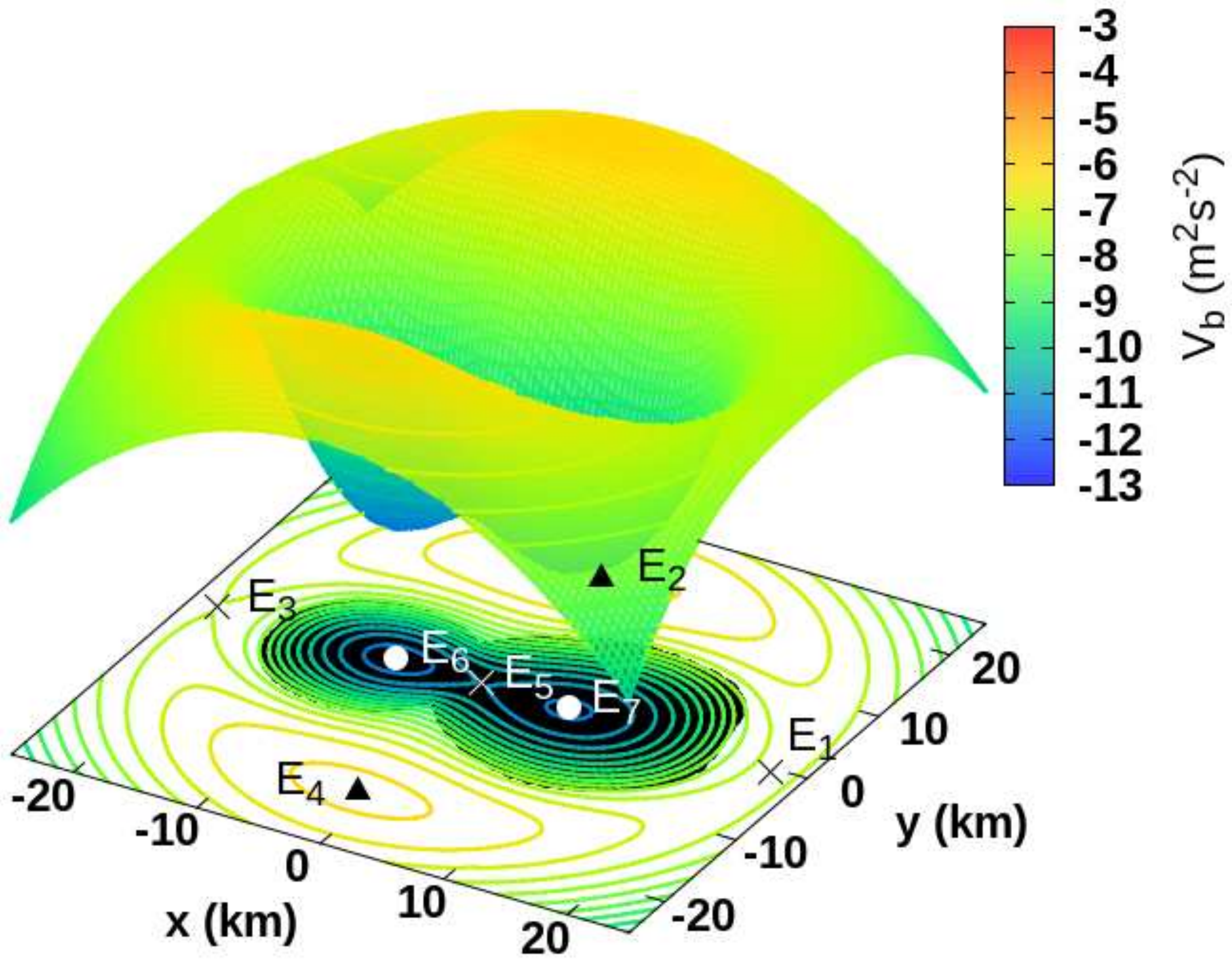}\\
  \includegraphics[width=8.6cm]{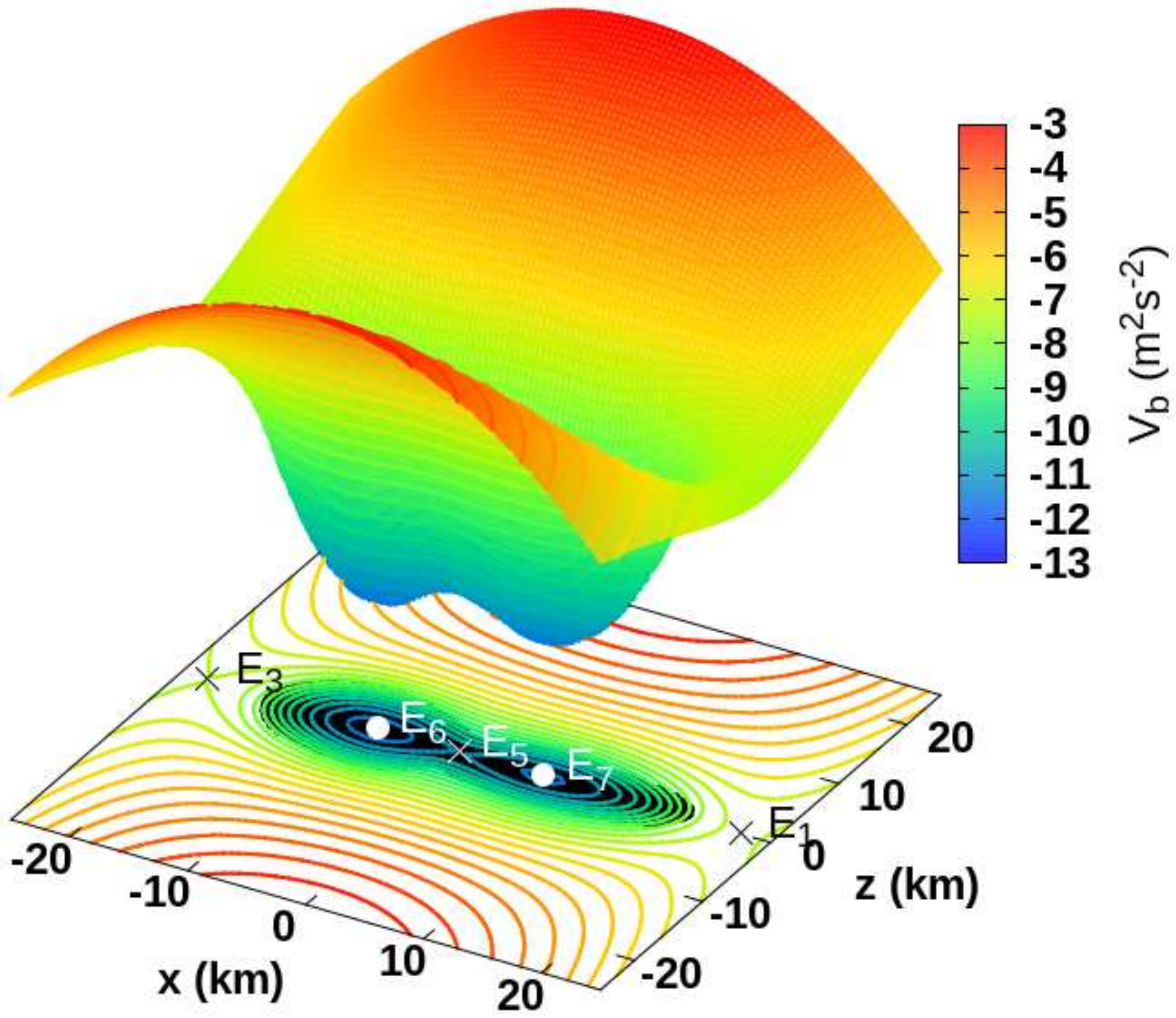}\\
  \includegraphics[width=8.6cm]{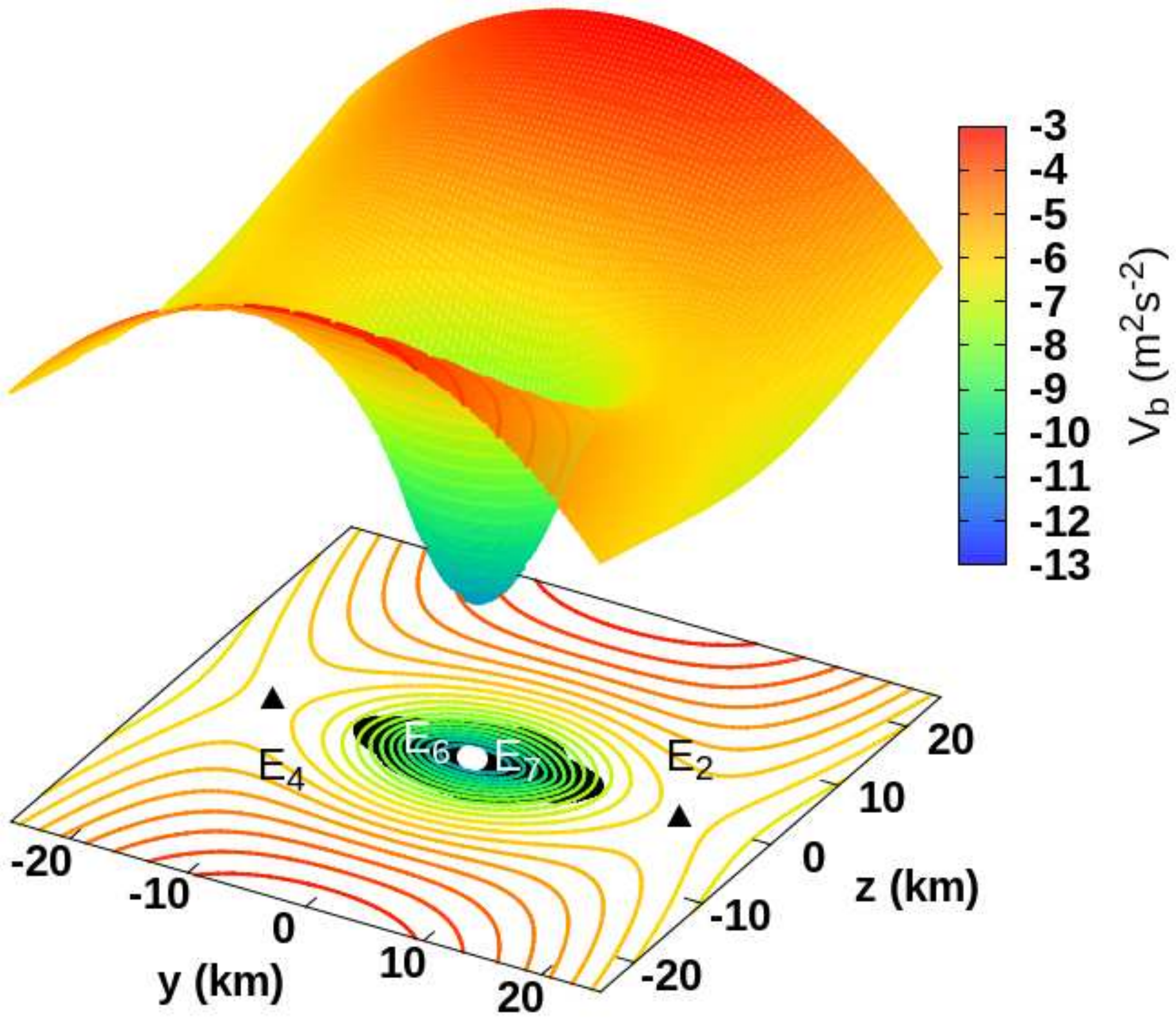}
  \caption{3-D plot of the binary geopotential $V_b(x,y,z)$ in the $xOy$, $xOz$ and $yOz$ planes, respectively. The color bar lines also illustrate zero-velocity curves of the binary Jacobi constant $J_b$, in m$^2$\,s$^{-2}$.}
  \label{fig:eq_8}
\end{figure} 
\begin{figure*}
   \centering 
   \fbox{\includegraphics[width=8.44cm]{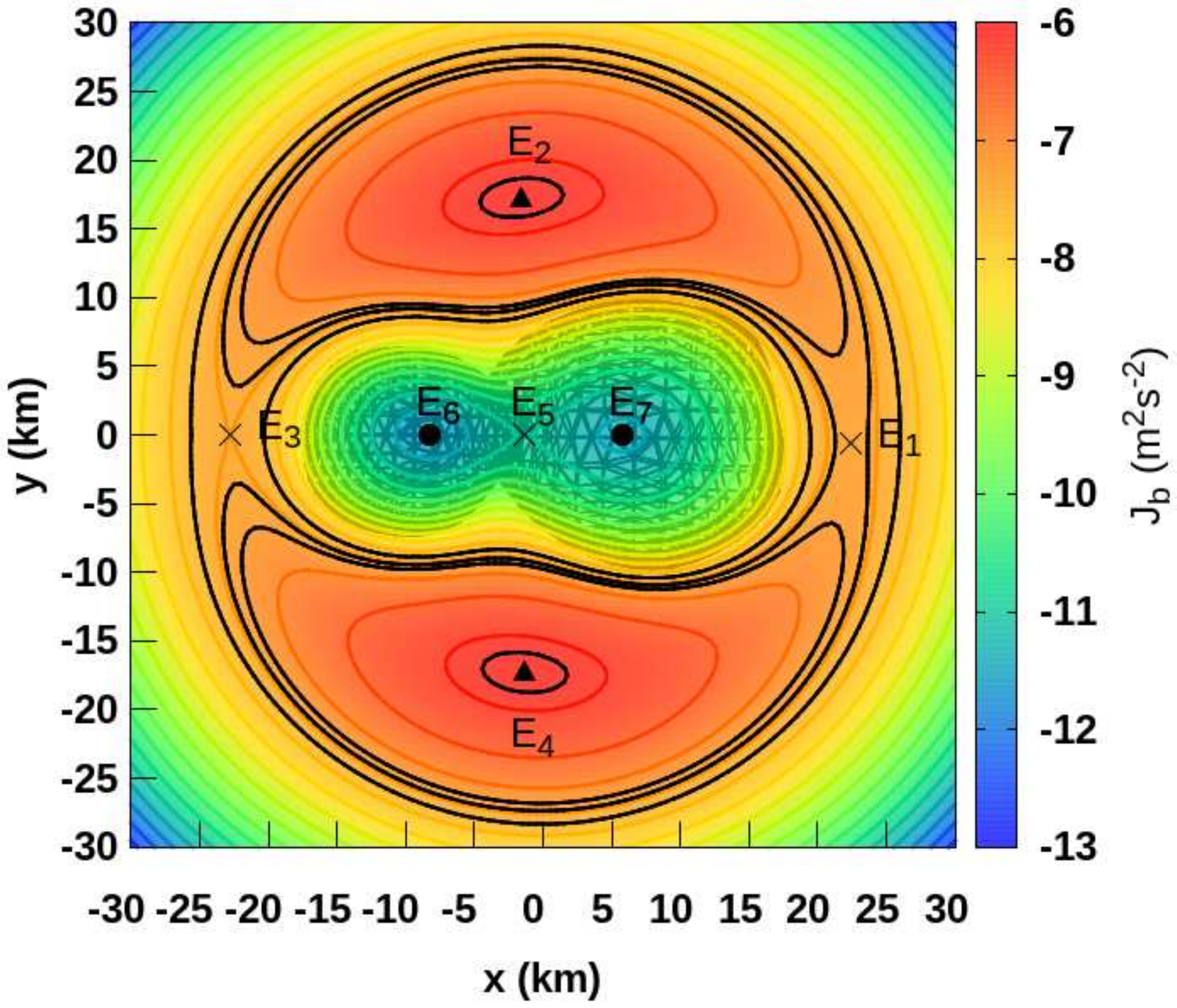}}
   \fbox{\includegraphics[width=8.44cm]{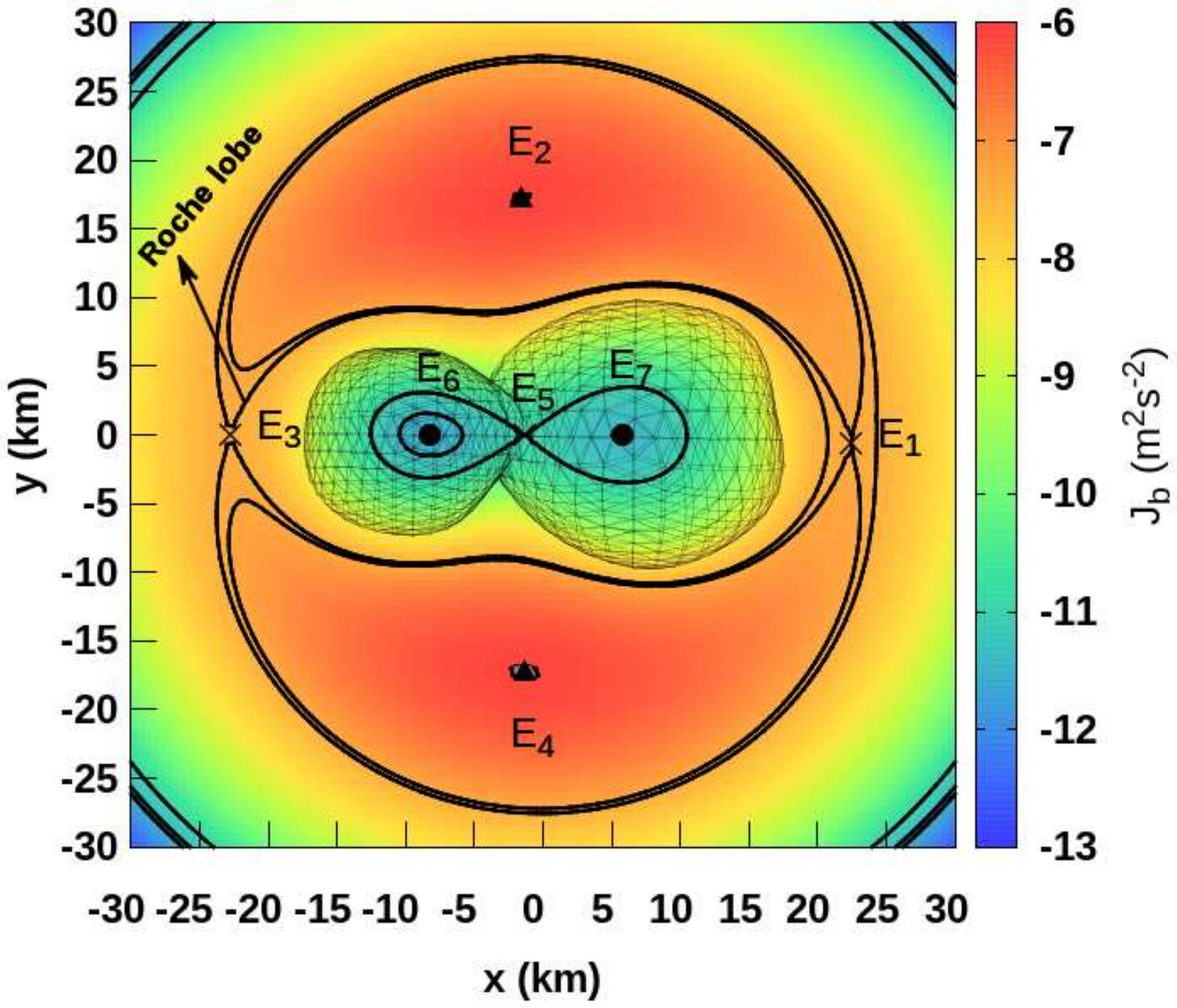}}
   \caption{(left-hand side) Zero-velocity curves in the equatorial plane of Arrokoth contact binary. Black lines represent each $J_b$ value. (right-hand side) Zero-velocity curves in the plane $xOy$. Black lines indicate zero-velocity-contour maps of each equilibrium point. The Roche-lobe $J_b^{\prime}$ is also indicated. Color box codes show the values of $J_b$, in m$^2$\,s$^{-2}$. In these figures, the shadowed areas sketch the shape of the Arrokoth.}
\label{fig:eq_7}
\end{figure*}
\begin{figure}
  \centering
  \fbox{\includegraphics[width=8.44cm]{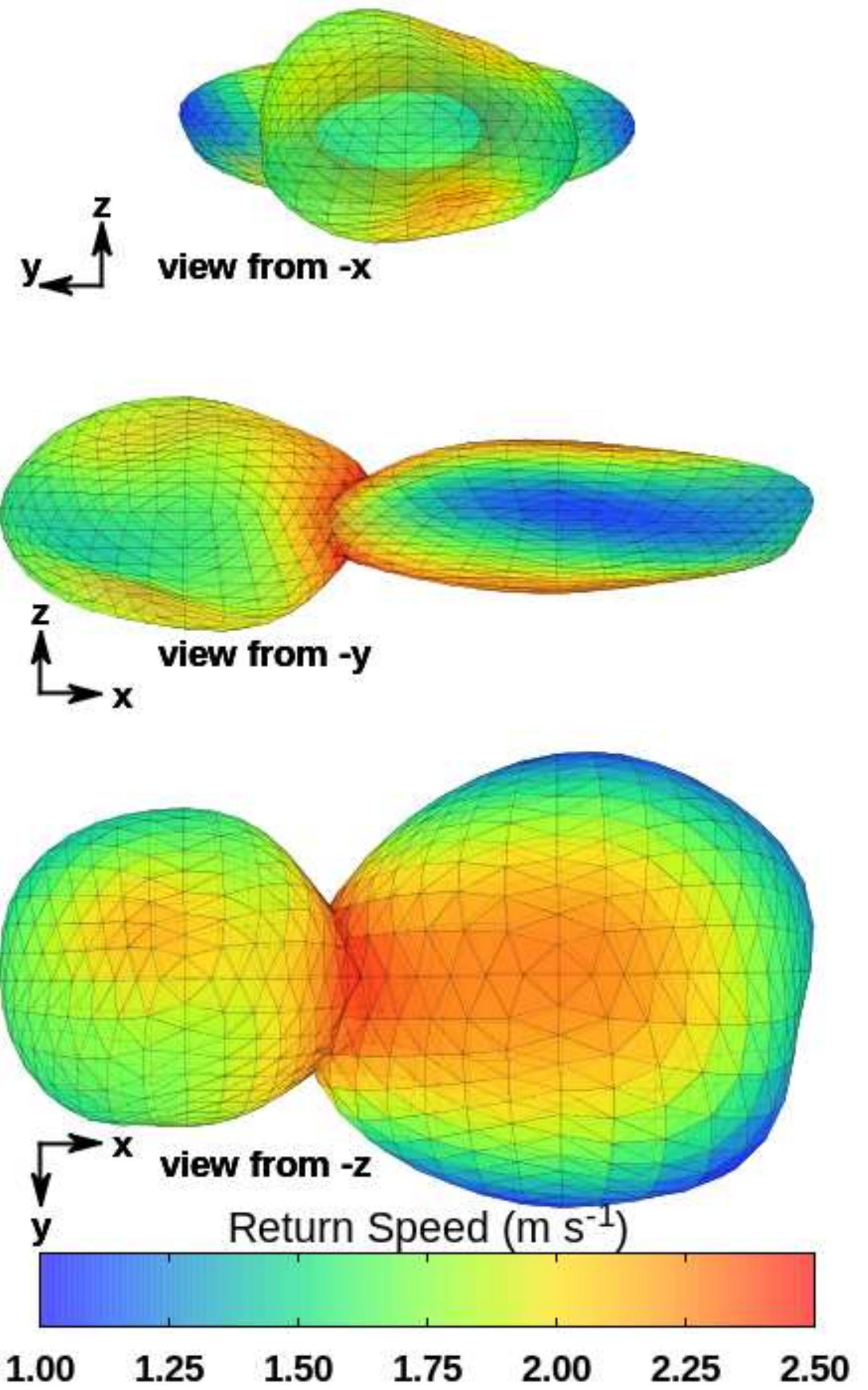}}
  \caption{Guaranteed return speed $v_r$ computed across the surface of Arrokoth contact binary, in m\,s$^{-1}$.}
  \label{fig:eq_6}
\end{figure}
\begin{figure*}
   \centering 
   \fbox{\includegraphics[width=8.44cm]{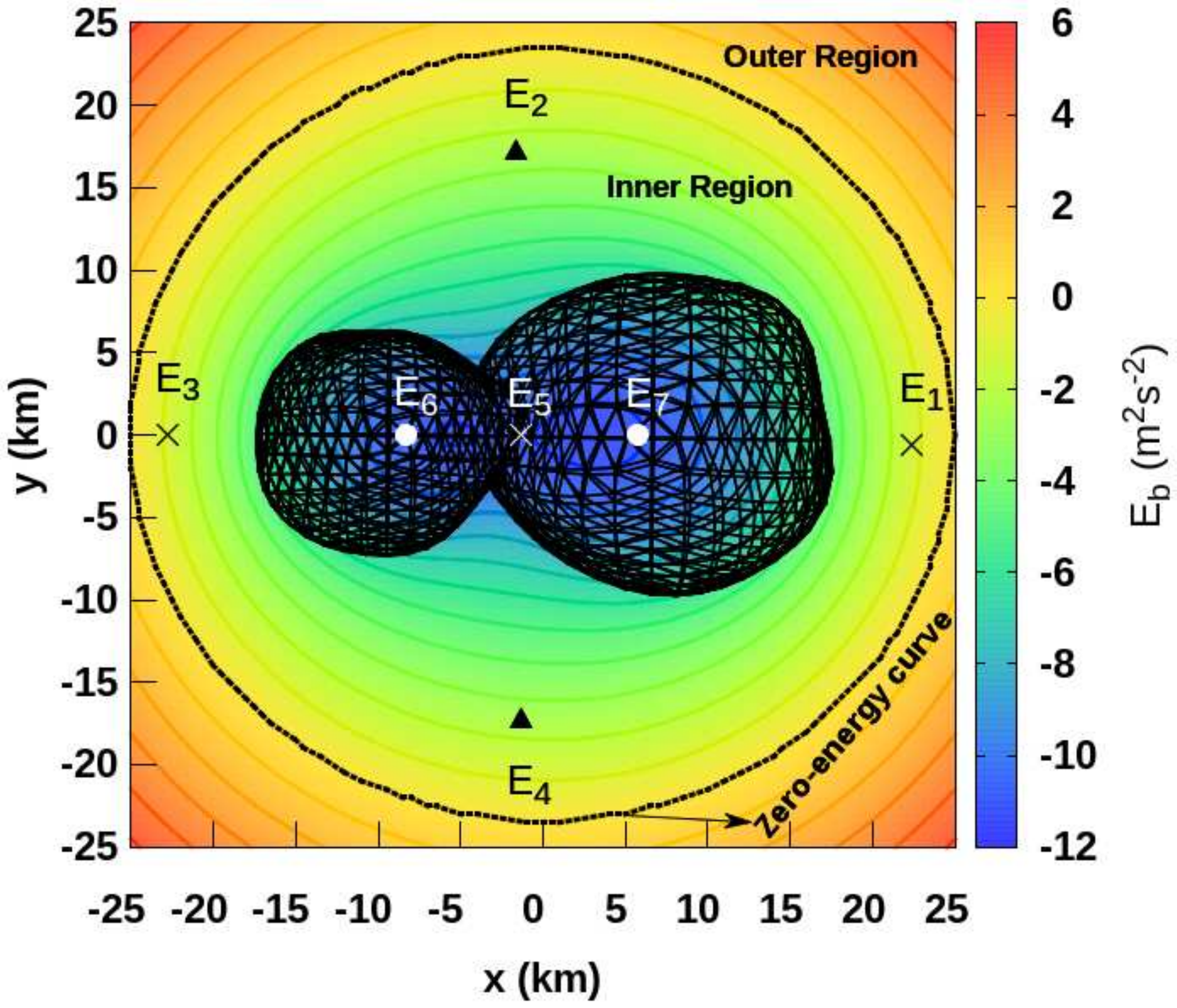}}
   \fbox{\includegraphics[width=8.44cm]{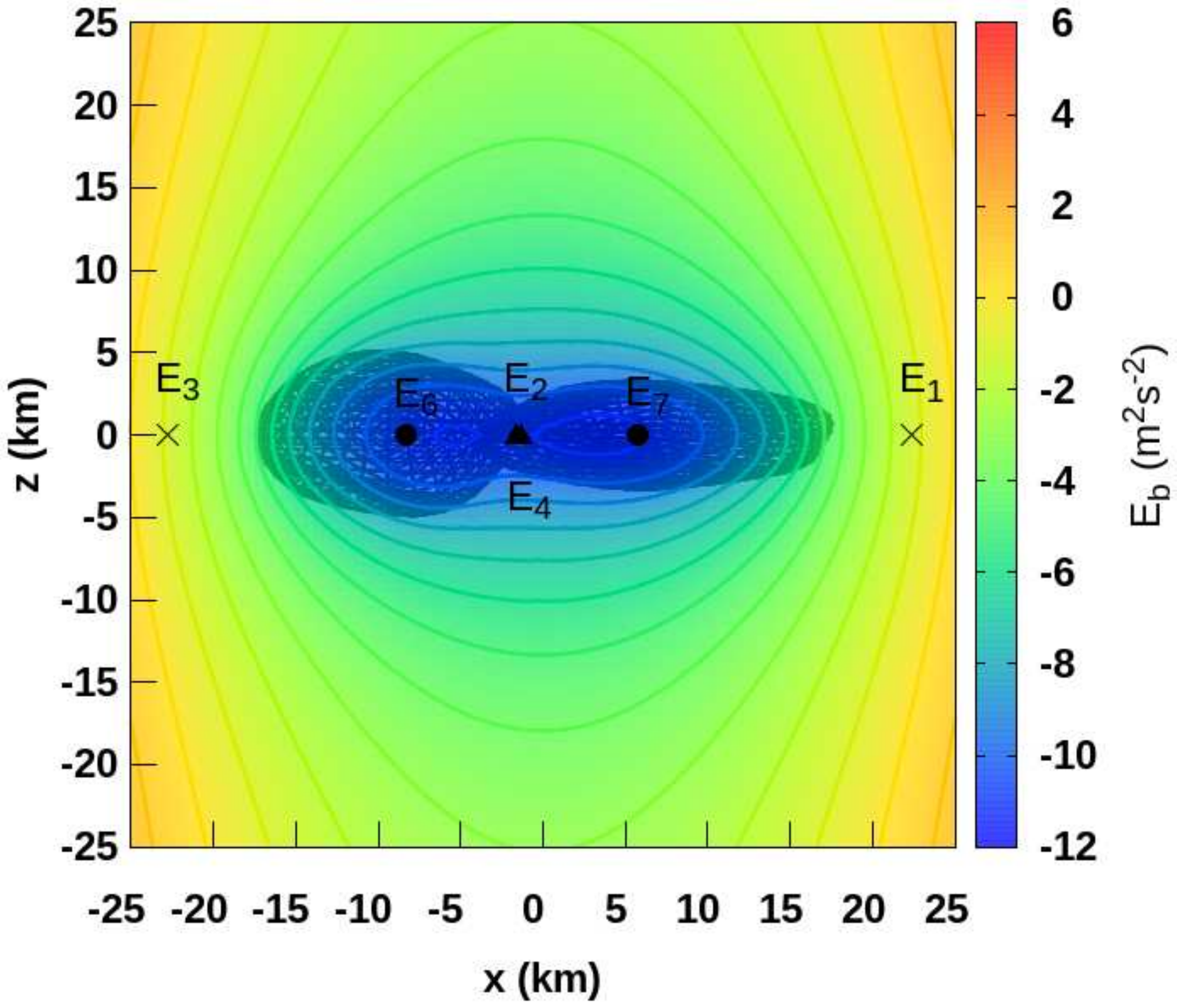}}
   \caption{Binary orbital energy lines $E_b$ in the projection planes $xOy$ and $xOz$, respectively. The black dashed line represents the zero-energy curve in the equatorial plane. Color bar codes denote the values of $E_b$, in m$^2$\,s$^{-2}$.}
\label{fig:eq_7b}
\end{figure*}
\begin{figure}
   \centering 
   \fbox{\includegraphics[width=8.44cm]{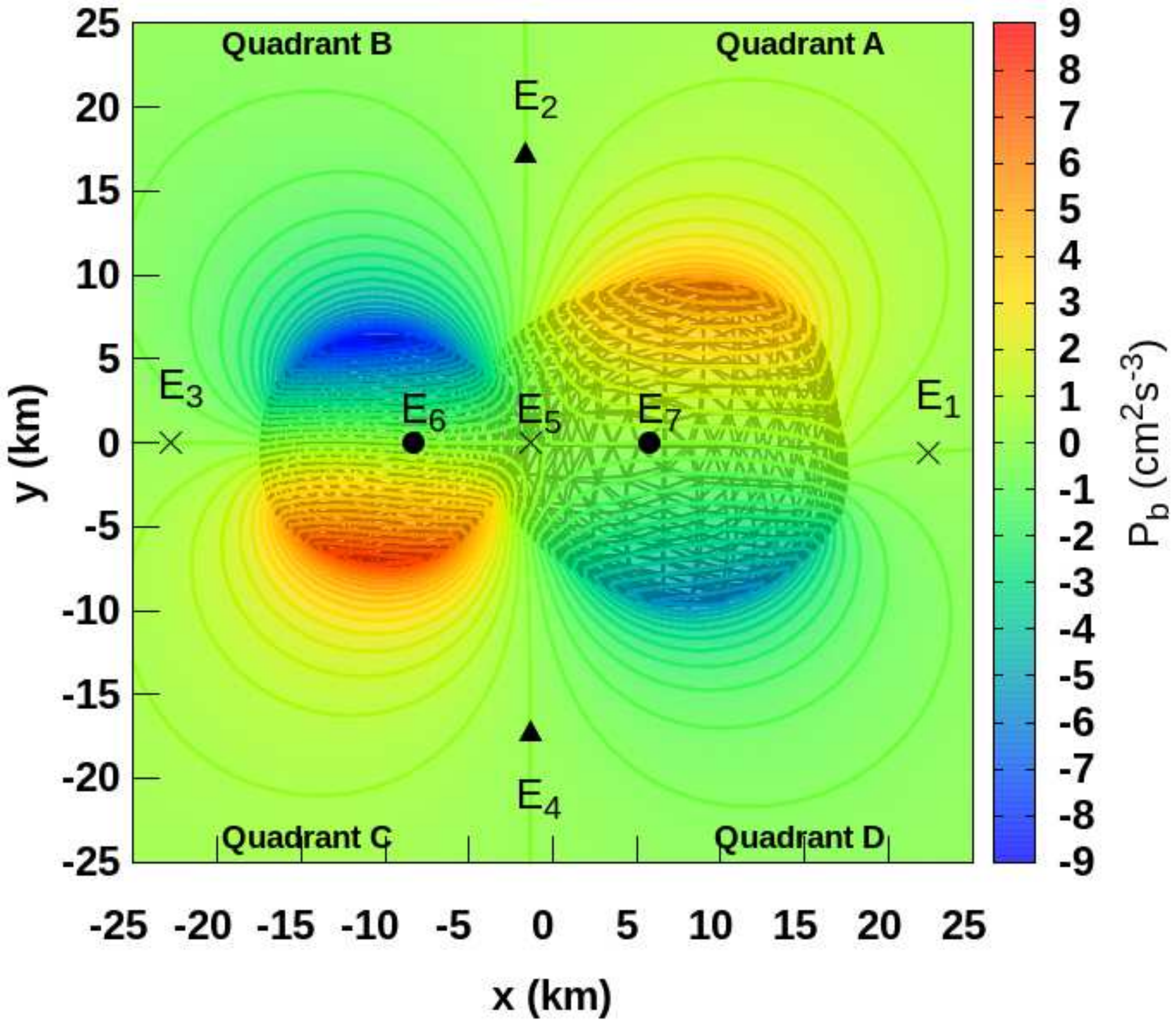}}
   \caption{The binary gravity power map $P_b$ in the equatorial plane. The color panels indicate the values of $P_b$, in cm$^2$\,s$^{-3}$. Shadowed areas sketch the shape of Arrokoth.}
\label{fig:eq_7c}
\end{figure}

The \textit{zero-velocity curves} provide insights into the dynamics of massless particles that may be ejected around Arrokoth from an impact or simple touchdown. Let us assume that Arrokoth contact binary has a uniform rotation about its largest moment of inertia ($z$-axis). Thus, the dynamic equations of motion of a test particle near the uniformly rotating Arrokoth in the binary body-fixed frame are \citep{Jiang2014b}:
\begin{eqnarray}
\label{eq:eq_7}
      \ddot{x} -2\omega \dot{y} +{{\partial V_b}\over{\partial x}} &=& 0, \nonumber \\
      \ddot{y} +2\omega \dot{x} +{{\partial V_b}\over{\partial y}} &=& 0, \\
      \ddot{z} +{{\partial V_b}\over{\partial z}} &=& 0. \nonumber
\end{eqnarray}

Because equations \eqref{eq:eq_7} are time-invariant, the binary Jacobi constant $J_b$ exists as an additional integral of motion. The Jacobi integral for the dynamic equations is:
\begin{eqnarray}
J_b &=& \underbrace{\frac{1}{2}(\dot{x}^2+\dot{y}^2+\dot{z}^2)}_{\text{kinetic energy}}\,\,\,\,\,\,\,\,\,\,\,\,\,+\underbrace{V_b(x,y,z)}_{\text{binary geopotential}}.
\label{eq:eq_8}
\end{eqnarray}
By analyzing Eq. \eqref{eq:eq_8} we can conclude that as the kinetic energy term is always positive, then
\begin{align}
J_b\geq V_b(x,y,z),
\label{eq:eq_9}
\end{align}
since the binary geopotential $V_b(x,y,z)$ is defined negatively. If we set the kinetic energy value equal to zero, then Eq. \eqref{eq:eq_8} yields:
\begin{align}
J_b &= V_b(x,y,z),
\label{eq:eq_10}
\end{align}
\noindent i.e., the binary geopotential can be used for a determined Jacobi integral value.

Thus, Eq. \eqref{eq:eq_10} provides constraints on the motion of a test particle and defines zero-velocity curves for the space, where a massless particle is allowed to be found and where it cannot be found, given a specific value for $J_b$. These results provide specific information about Arrokoth Hill's stability \citep{Murray2000}, i.e., concerning the possible movement of a test particle around Arrokoth.
Figure \ref{fig:eq_8} presents Arrokoth's binary geopotential in different planes. The colour contour maps denote the binary Jacobi constant values $J_b$. The projection planes confirm that Arrokoth contact binary has a total of seven equilibrium points. Additionally, these five equilibrium points are near the equatorial plane. Similarly, there are three equilibrium points $E_1$, $E_3$ and $E_5$ close to the $xOz$ plane. Because all of the equilibrium points are in the vicinity of Arrokoth contact binary's equatorial plane, we can use the structure of the zero-velocity curves and the projections of the equilibrium points to judge their stability. From the zero-velocity-contour maps, we can see that there are no stable locations in the proximity of the equilibrium points $E_1$, $E_3$ and $E_5$. All three projection planes show saddle-like structures around these points, which is compatible with their linear stability analysis (Table \ref{tab:eq_2}). However, for the equilibrium points $E_2$, $E_4$, $E_6$ and $E_7$, there are centre-like structures around their locations in at least one projection plane. This occurs in the $xOy$ plane for the four external points and also occurs in the $xOz$ plane for the $E_6$ and $E_7$ points.

\subsection{Return Speed}
\label{sec:eq:rspeed}
From the Roche lobe concept, we can find the guaranteed \textit{return speed} across the Arrokoth lobes' surfaces using the velocity from the kinetic energy term from Eq. \eqref{eq:eq_8} as $v_r$ \citep{Scheeres2012}. Thus, we have:
\begin{align}
v_r = \sqrt{2(J_b^{\prime}-V_b(x,y,z))},
\label{eq:eq_11}
\end{align}
\noindent where $J_b^{\prime}$ is the value of the Jacobi constant for the Roche lobe thresholds and whose $v_r$ value is set to zero, if the binary geopotential $V_b(x,y,z)$ across the surface of the body exceeds $J_b^{\prime}$.

The right-hand side of Figure \ref{fig:eq_7} shows the Roche lobe boundaries in the equatorial plane made using the Arrokoth binary geopotential value $V_b(x,y,z)$ at equilibrium point $E_3$ (black line). In other words, the zero-velocity curve value of $J_b^{\prime}=-7.40$\,m$^2$\,s$^{-2}$ connects the inner and outer branches through equilibrium point $E_3$ (i.e., the least unstable external equilibrium point, see \ref{tab:eq_1}). From that binary Jacobi constant value, we derived an upper value for Arrokoth's surface ejecta speed, which ensures that all ejecta with speeds less than this value will eventually fall back onto Arrokoth's surface. In addition, in the right-hand side of Fig. \ref{fig:eq_7} we also show the zero-velocity curves from the other equilibrium points (black lines). Note that the three innermost and the three outermost (out-of-graph) black lines are due to the zero-velocity-contour map from internal equilibrium points $E_5$, $E_6$ and $E_7$ (note that the $E_6$ and $E_7$ zero-velocity curves are asymmetric and very close to each other). From Eq. \eqref{eq:eq_11}, if $V_b(x,y,z)\geq J_b^{\prime}$, then $v_r=0$, and a test particle can escape from the environment near Arrokoth, even though it is on its surface. Otherwise, if $V_b(x,y,z)< J_b^{\prime}$, and considering that a test particle is positioned within the zero-velocity curve that surrounds Arrokoth contact binary, there is insufficient energy for it to escape from the system. Therefore, the particle will return and it will collide with one of the lobe surfaces. Figure \ref{fig:eq_6} shows the guaranteed return speed $v_r$ computed over the surfaces of the large and small lobes, in m\,s$^{-1}$. Arrokoth's limiting speeds are larger at the polar regions of the lobes and in the neck, while the lowest values of the guaranteed return speed are concentrated in the equatorial region.

\subsection{Binary Orbital Energy}
We can also use the behaviour of the \textit{binary orbital energy} combined with zero-velocity curves and return speeds to infer the boundaries of the captured and escaped orbits of a test particle in the environment around Arrokoth. Thus, the binary orbital energy can be written as \citep{Scheeres1996}:
\begin{eqnarray}
E_b &=& \underbrace{\frac{1}{2}(\dot{\textbf{r}}+\pmb{\Omega}\times\textbf{r})\cdot(\dot{\textbf{r}}+\pmb{\Omega}\times\textbf{r})}_{\text{rotational potential}}\,\,\,\,\,\,\,\,\,\,\,\,\,+\underbrace{U_b(\textbf{r})}_{\text{gravitational potential}}.
\label{eq:eq_12}
\end{eqnarray}
\noindent Note from Eq.\eqref{eq:eq_12} that the binary orbital energy is relative to the inertial frame and is represented by the quantities of the binary body-fixed frame with constant vector $\pmb{\Omega}$ in the $z$-axis direction.

We have a captured orbit for $E_b < 0$ and an escaped orbit for $E_b > 0$. Combining the scalar form of the binary Jacobi constant (Eq. \eqref{eq:eq_8}) and the return speed (Eq. \eqref{eq:eq_11}) into the binary orbital energy equation, then Eq. \ref{eq:eq_12} yields: 
\begin{align}
E_b = J_b+\omega^2 r(x,y)^2+\delta\omega v_r r(x,y),
\label{eq:eq_12b}
\end{align}
\noindent where the last item represents the inner product of the convected velocity vector ($\pmb{\Omega}\times\textbf{r}$) and the relative velocity vector, $\delta$ indicates the cosine of the angle between these two vectors and $r(x,y)=\sqrt{x^2+y^2}$ represents the intensity of the radius vector on the equatorial plane $xOy$.

Equation \ref{eq:eq_12b} shows that the magnitude of the binary orbital energy $E_b$ depends on the binary Jacobi integral $J_b$ and the return speed $v_r$, which provide a direct criterion to distinguish captured and escaped orbits in the neighbourhood of Arrokoth. Thus, let us set $v_r=0$ for a possible test particle escape orbit from the environment near Arrokoth. Then, Eq. \ref{eq:eq_12b}, can be written as:
\begin{align}
E_b = \frac{1}{2}\omega^2(x^2+y^2)+U_b.
\label{eq:eq_12c}
\end{align}

This can also be considered as the binary orbital energy of a massless particle on the zero-velocity curve. Thus, on the left-hand side of Fig. \ref{fig:eq_7b}, we show the binary orbital energy on Arrokoth's zero-velocity curves. Additionally, the zero-energy curves can be drawn numerically in the first-order approximation of binary gravitational force potential $U_b$, if we set $E_b = 0$. Thus the Eq. \eqref{eq:eq_12c} in the equatorial plane ($z=0$) becomes:
\begin{align}
r(x,y) & = \sqrt{\dfrac{-2U_b(x,y)}{\omega^2}}.
\label{eq:eq_13}
\end{align}

As shown in the left-hand side of Fig. \ref{fig:eq_7b}, the zero-energy curve relative to Arrokoth contact binary has an average radius of 24.14\,km. We deduce that the binary gravitational force potential $U_b$ is nearly uniform at this distance from Arrokoth's centre mass. This result reflects that Arrokoth's shape irregularity only has an evident influence on the gravitational field in a nearby area, but weak influence in the area at a certain distance from its barycentre. This is a perspective from the equatorial plane $xOy$. The right-hand side of Fig. \ref{fig:eq_7b} shows the binary energy from the projection plane $xOz$. We note that the equipower curves in this case extend to infinity in the $\pm z$-axes direction and shrink around Arrokoth contact binary. Then, a massless particle is more likely to be captured in equatorial orbits than polar orbits. The zero-energy curve divides the zero-velocity curves equatorial plane into two regions. We call the inner area the `inner region' where $E_b<0$, as shown in the left-hand side of Fig. \ref{fig:eq_7b}. A test particle with an orbit inside the inner region must be in a captured orbit. We call the outer area the `outer region', where $E_b>0$. A massless particle with an orbit in the outer region should escape from the system. All seven equilibrium points of Arrokoth contact binary are located in the inner region of the zero-energy curve.

\subsection{Binary Gravity-Power}
\label{sec:eq:pow}
Finally, combined with the Jacobi integral, the binary orbital energy Eq. \ref{eq:eq_12} can be rewritten as \citep{Yu2013}:
\begin{align}
E_b = J + \Omega\cdot\pmb{L},
\label{eq:eq_12d}
\end{align}
\noindent where $\pmb{L} = \textbf{r}\times(\dot{\textbf{r}}+\pmb{\Omega}\times\textbf{r})$ is the massless particle's angular orbital momentum.

Taking the derivative of Eq. \ref{eq:eq_12d} with respect to time, the formula of \textit{binary gravity-power} can be obtained as:
\begin{align}
P_b & = \omega\bigg(x{{\partial U_b}\over{\partial y}}-y{{\partial U_b}\over{\partial x}}\bigg).
\label{eq:eq_14}
\end{align}

Eq. \eqref{eq:eq_14} is only position-dependent. The binary potential is fully determined by the geometry of the gravitational field, which is useful to measure increases and decreases in the binary orbital energy. The left-hand side of Fig. \ref{fig:eq_7c} illustrated the gravity-power field of Arrokoth contact binary, which divided the equatorial plane $xOy$ into four quadrants: `A' ($x>0$ and $y>0$), `B' ($x<0$ and $y>0$), `C' ($x<0$ and $y<0$) and `D' ($x>0$ and $y<0$), respectively. We also show the location of all seven equilibrium points and the shadowed area sketches the shape of each lobe. In quadrants A and C, the binary gravity-power equation is defined positively, i.e., $P_b>0$. However, in quadrants B and D, the binary gravity-power field is less than zero ($P_b<0$). Therefore, a massless particle has the binary orbital energy $E_b$ added in the regions where $P_b>0$, but $E_b$ decreases in the areas where $P_b<0$. The positive and negative zones both account for a $\sim 50\%$ area of the plane. In addition, the equilibrium points in the equatorial plane lie in the binary zero-gravity power curves, which are locations where $P_b=0$. For example, a massless particle that surrounds the equilibrium point $E_2$ in a periodic orbit near-equatorial plane has half of its trajectory in quadrant A, while another half of its path is in quadrant B. Then, the test particle increases its binary orbital energy in quadrant A and decreases in quadrant B, i.e., $E_b$ changes periodically over time.

Figure \ref{fig:eq_7c} indicates that extreme power values are reached at locations where the terrain becomes significantly steeper. The figure suggests more possibilities for the surface particles in quadrants B and D to be ejected from the large and small lobes than those of quadrants A and C, which has implications for the regolith evolution, which is related to both the dynamics of ejected particles and the topography of minor bodies \citep{Scheeres2002}. 

\section{Final Comments}
\label{sec:final}
This study provided insights into the exploration of the surface dynamics, equilibrium points, and individual lobes of the New Horizons' targeted Kuiper Belt object (486958) Arrokoth contact binary. Firstly, we produced a low facet polyhedral model of Arrokoth using $1,046$ vertices and $2,928$ edges combined into $1,952$ triangular faces. Their geometric and physical features were also explored using the concept of geometric height. The surface orientation of the large and small lobes was low and Arrokoth's surface tilts did not exceed $90^\circ$. 

We computed Arrokoth's binary gravitational force potential using our mathematical approach. The binary geopotential allowed us to study the dynamics of the surface environment through several quantities. Our results show that the equatorial regions of the large and small lobes are binary geopotential highs keeping the surface accelerations between $0.5$ and $1$\,mm\,s$^{-2}$. If the dynamic slope angles are $<40^\circ$, then loose particles below the friction angle can be trapped in depression sites, like some craters found across Arrokoth's surface. We found that the equatorial area of the large lobe is an unstable region, while the poles concentrate the flow tendency of surface particles along with the neck zone, in contrast with most of the minor bodies, which have a small spin period. The results suggest that Arrokoth's polar areas can retain some free particles. In addition, the overall picture of the surface slope angles and of the tangential acceleration vector fields do not vary significantly for densities up to 0.25\,g\,cm$^{-3}$. The escape speed across Arrokoth's surface lies between $2.5-8.5$\,m\,s$^{-1}$, and in its neck, the escape speeds can achieve $7.9$\,m\,s$^{-1}$.

Next, we computed the equilibrium points in Arrokoth contact binary's gravitational field. In addition, we also found the equilibrium points for each lobe by considering their gravitational fields separately. We found seven equilibrium points for Arrokoth contact binary. All external equilibrium points have no radial symmetry. Arrokoth's zero-velocity curves show saddle-like structures around equilibrium points $E_1$, $E_3$ and $E_5$, which is compatible with the linear stability analysis. However, centre-like structures appear around equilibrium points $E_2$, $E_4$, $E_6$ and $E_7$ in at least one projection plane. The inner equilibrium point $E_5$ is unstable, which shows the instability of the neck. The neck is distant from Arrokoth's barycentre by $\sim 1.37$\,km. Meanwhile, the outer equilibrium point $E_4$ is the most unstable. Moreover, the large and small lobes have five equilibrium points with different topological structures from those found in Arrokoth. We also explored the effects of upper and lower densities on the dynamic properties of equilibrium points. We found that when upper densities are considered, the external equilibria move far away from Arrokoth, and considering lower density values, the external equilibrium points move towards Arrokoth until some of them reach its surface and vanish. Arrokoth's stability through 1:1 resonance was also investigated. Our analysis suggests that, in some areas, the body will be in tension and materials will be thrown off from Arrokoth's surface. Additionally, at equilibria distance and beyond, the second-order and degree gravitational force potential can be used as a good approximation for Arrokoth contact binary's real gravitational field. We also studied Arrokoth Hill's stability through zero-velocity curves and guaranteed return speeds. Arrokoth's Roche lobe is located around $-7.40$\,m$^2$\,s$^{-2}$ of the binary Jacobi constant. The guaranteed return speed thresholds are higher in the polar areas of the lobes and in the neck region, and lower in the equatorial regions. Finally, the Arrokoth contact binary has a peculiar binary gravity-power field that differs from other minor bodies. Arrokoth's binary energy increases in quadrants A and C and decreases in quadrants B and C, which is converse to other prolate minor bodies.

\section*{Acknowledgements}
The authors thank an anonymous reviewer whose comments greatly improved the manuscript, Improvement Coordination Higher Education Personnel - Brazil (CAPES) - Financing Code 001 and National Council for Scientific and Technological Development (CNPq, proc. 305210/2018-1). The research also had computational resources provided by the thematic project FAPESP (proc. 2016/24561-0) and the Center for Mathematical Sciences Applied to Industry (CeMEAI), funded by FAPESP (grant 2013/07375-0). We are also grateful to the entire New Horizons team for making the encounter with KBO Arrokoth possible.

\section*{ORCID IDS}
A. Amarante \href{https://orcid.org/0000-0002-9448-141X}{\includegraphics[scale=0.5]{orcid_16x16.pdf}} \url{https://orcid.org/0000-0002-9448-141X}\\
O. C. Winter \href{https://orcid.org/0000-0002-4901-3289}{\includegraphics[scale=0.5]{orcid_16x16.pdf}} \url{https://orcid.org/0000-0002-4901-3289}

\section*{Data availability}
The data that support the findings of this study are available from the corresponding authors upon reasonable request.

\section*{Code availability}
Simulation codes used to generate these results and generated data are available online at \url{https://github.com/a-amarante}.




\bibliographystyle{mnras}
\bibliography{biblio} 



\appendix



\section{Computation of the Binary Gravitational Field}
\label{sec:app}



The closed expressions with singularities correction terms for the gravitational force potential, the gravity attraction vector and the gravity gradient matrix, are respectively:
\begin{equation}
\begin{split}
U_k(x_1,x_2,x_3) &= -\frac{G\rho}{2}\sum_{p=1}^n\sigma_ph_p\bigg[\sum_{q=1}^m\sigma_{pq} h_{pq}LN_{pq}\\&
+h_p\sum_{q=1}^m\sigma_{pq}AN_{pq}+\sin(g_{\mathcal{A}_p})\bigg],
\end{split}
\label{eq:grav_01}
\end{equation}
\begin{equation}
\begin{split}
-\pdv{U_k(x_1,x_2,x_3)}{x_i} &= -G\rho\sum_{p=1}^n\cos{(\textbf{N}_p,\textbf{e}_i)}\bigg[\sum_{q=1}^m\sigma_{pq}h_{pq}LN_{pq}\\&
+h_p\sum_{q=1}^m\sigma_{pq}AN_{pq}+\sin(g_{\mathcal{A}_p})\bigg]\,\,\,\,\,\, (i= 1,2,3),
\end{split}
\label{eq:grav_2}
\end{equation}
\begin{equation}
\begin{split}
-\pdv{U_k(x_1,x_2,x_3)}{x_i}{x_j} &= G\rho\sum_{p=1}^n\cos{(\textbf{N}_p,\textbf{e}_i)}\bigg[\sum_{q=1}^m\cos{(\textbf{n}_{pq},\textbf{e}_j)}LN_{pq}\\&
+\sigma_p\cos{(\textbf{N}_p,\textbf{e}_j)}\sum_{q=1}^m\sigma_{pq}AN_{pq}+\sin(g_{\mathcal{B}_{pj}})\bigg]\,\\& \,\,\,\,\,\,\,\,\,\,\,\,\,\,\,\,\,\,\,\,\,\,\,\,\,\,\,\,\,\,\,\,\,\,\,\,\,\,\,\,\,\,\,\,\,\,\,\,\,\,\,\,\,\,\,\,\,\,\,\,\,\,\,\,\,\,\,\,\,\,\,\,\,\,\,\,\,\,\,\,\,\,\,\,\,\,\,\,\,\,\,\,\,\,\,\,\,\,\,\,\,\,\,\,\,\,\,\,\,\,\,\,\,\, (i,j = 1,2,3);
\label{eq:grav_3}
\end{split}
\end{equation}
\noindent where
\begin{eqnarray}
LN_{pq}= \ln(\frac{s_{2pq}+l_{2pq}}{s_{1pq}+l_{1pq}}), \nonumber \\ \label{eq:grav_4}
 \\
AN_{pq}= \arctan(\frac{h_ps_{2pq}}{h_{pq}l_{2pq}})-\arctan(\frac{h_ps_{1pq}}{h_{pq}l_{1pq}}). \nonumber
\end{eqnarray}

\begin{figure}
  \centering
  \fbox{\includegraphics[width=\linewidth]{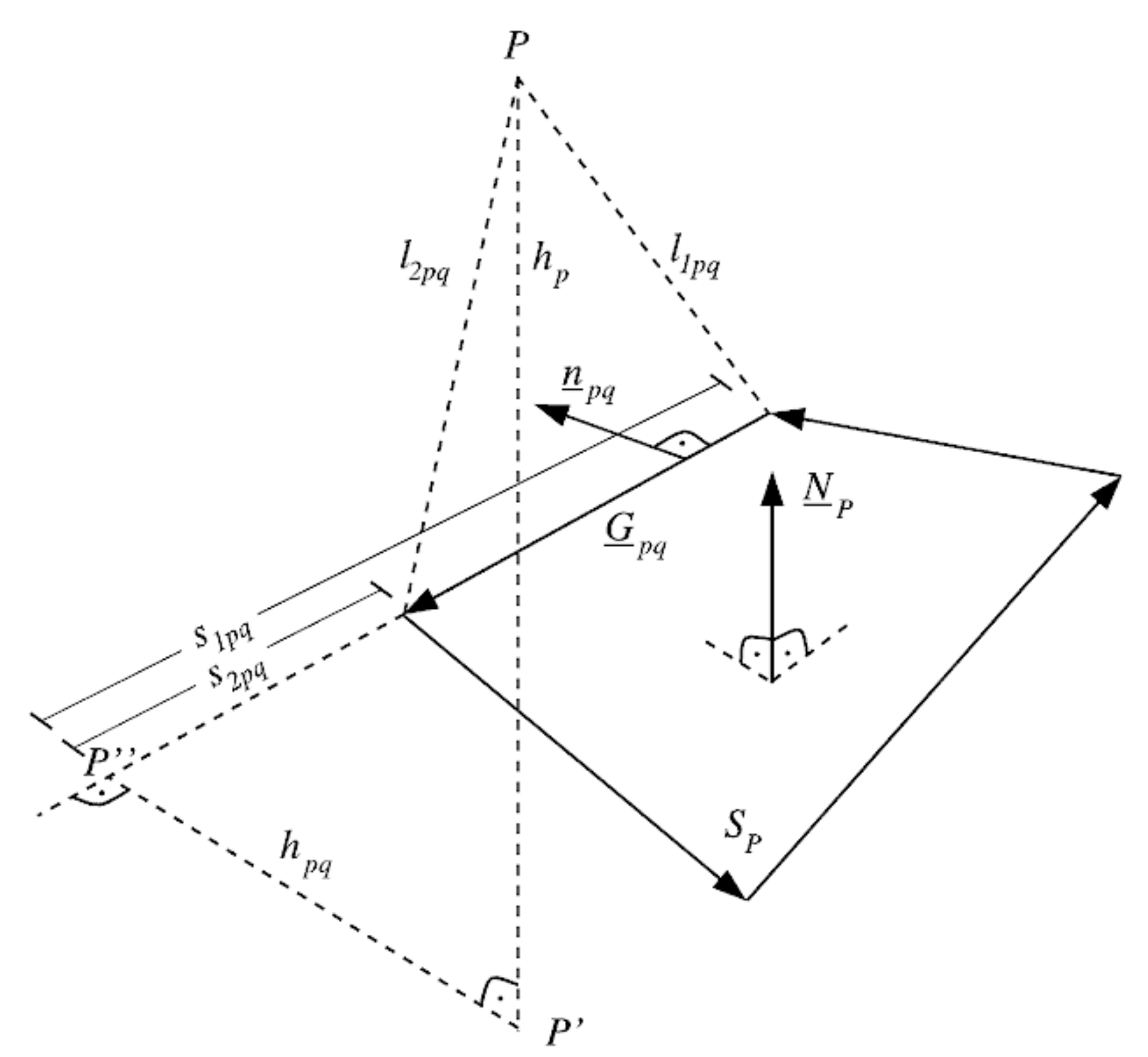}}
  \caption{Geometrical meaning of the quantities used to represent the gravitational force potential and its derivatives of the polyhedral model \citep{Tsoulis2001}.}
  \label{fig:grav_1}
\end{figure}

\noindent \textbf{where, $x_1$, $x_2$ and $x_3$ are the coordinates of a test particle.} In Figure \ref{fig:grav_1} we have displayed all geometrical quantities that appearing from Eqs. \eqref{eq:grav_01}-\eqref{eq:grav_4}. The polyhedron is assumed to be homogeneous with constant volume density $\rho=M/V$, with $n$ faces, each having $m$ sides, where $G=6.67408 \times 10^{-20}$\,km$^3$\,kg$^{-1}$\,s$^{-2}$ is the gravitational constant\footnote{CODATA - \url{http://physics.nist.gov/constants}}. Subscript $k=1,2$ refers to the single gravitational force potential from lobes large and small, respectively. And subscript $i=1,2,3$ denotes in the binary body-fixed coordinate frame each coordinate axis $x$, $y$ and $z$, respectively. Each face defines a plane that is represented by polygonal surface $S_p$ and it have the normal vector \textbf{N}$_p$. The orthogonal projection of point $P(x_1,x_2,x_3)$ on the plane of the polygon $S_p$ is denoted by $P'$ and $P''$ is the orthogonal projection of $P'$ on the straight line defined by the segment $G_{pq}$. The distance between points $P$ and $P'$ is denoted by $h_p$ and the distance between points $P'$ and $P''$ is represented by $h_{pq}$. $\textbf{n}_{pq}$ is the unit vector which belongs to the plane of the polygon $S_p$ and it is pointing per definition outside the closed polygonal surface $S_p$. If $\sigma_{pq}=-1$, then $\textbf{n}_{pq}$ points to the half-plane containing the point $P'$ and otherwise $\sigma_{pq}=+1$ if it points to the other half-plane. $\cos{(\textbf{N}_{p},\textbf{e}_t)}$ and $\cos{(\textbf{n}_{pq},\textbf{e}_t)}$ denote the direction cosines between normal vectors $\textbf{N}_{p},\textbf{n}_{pq}$ and unit vectors basis $\textbf{e}_t$ ($t=1,2,3$), respectively. Since $|\textbf{N}_p|=|\textbf{n}_{pq}|=1$, then we can compute the direction cosines as $\cos{(\textbf{N}_{p},\textbf{e}_k)}=\textbf{N}_p\cdot \textbf{e}_k$ and $\cos{(\textbf{n}_{pq},\textbf{e}_k)}=\textbf{n}_{pq}\cdot \textbf{e}_k$. $l_{1pq}$ and $l_{2pq}$ are the $3$-D distances between $P$ and the end points of $G_{pq}$. $s_{1pq}$ and $s_{2pq}$ denote the $1$-D distances between $P''$ and the two end points of segment $G_{pq}$, respectively. Terms $LN_{pq}$ and $AN_{pq}$ are abbreviations of the transcendental functions given by Eqs. \eqref{eq:grav_4}. Finally, the terms $\sin (g_{\mathcal{A}p})$ and $\sin (g_{\mathcal{B}_{pj}})$ are the singularity terms that appear for specific locations of $P'$ with respect to the polygonal line $G_p$ when one attempts to apply the Gauss divergence theorem for these cases. \citet{Tsoulis1999} and \citet{Tsoulis2001} showed the values of the three singularity cases: when $P'$ lie inside $S_p$, $P'$ is located on segment $G_p$, but does not at any of its vertices, and $P'$ is located at one of $G_p$'s vertices. When $P'$ is located outside $S_p$, then singularity terms vanishes: $\sin (g_{\mathcal{A}p})=\sin (g_{\mathcal{B}_{pj}})=0$. From polyhedra approach Eqs. \eqref{eq:grav_01}-\eqref{eq:grav_4} we can computed the binary gravitational force potential, the binary gravity attraction and the binary gravity gradient matrix as a summation: $U_b=\sum_{k=1}^2U_k$, $-\nabla U_b=\sum_{k=1}^2\bigg(-\pdv{U_k}{x_i}\bigg)$ and $-\nabla \nabla U_b=\sum_{k=1}^2\bigg(-\pdv{U_k}{x_i}{x_j}\bigg)$ with $i,j=1,2,3$ and where $\nabla$ represents the Hamiltonian operator.

\section{Eigenvalues}
\label{sec:eigen}
Eigenvalues of the equilibrium points presented in Table \ref{tab:eq_1}, referring to the (486958) Arrokoth contact binary system.
\begin{table}
 \centering
  \caption{Eigenvalues ($\gamma_n\times 10^{-4}$, $n=1,2,...,6$) of equilibrium points in the gravitational field of the Arrokoth, large and small lobes with their topological structures. They are computed for a uniform density of $\rho=0.5$\,g\,cm$^{-3}$.}
 \label{tab:eq_2}
 \scalebox{0.9}
 {
 \begin{tabular}{ccccc}
  \toprule
  Point & $\gamma_{1,2}$ & $\gamma_{3,4}$ & $\gamma_{5,6}$ & Topological Structure \\
  \hline
  \multicolumn{5}{c}{Arrokoth} \\
  $E_1$ & $\pm1.195$ & $\pm1.498i$ & $\pm1.259i$ & saddle--centre--centre \\
  $E_2$ & $-0.669 \pm0.990i$ & $0.669 \pm0.990i$ & $\pm1.156i$ & sink--source--centre \\
  $E_3$ & $\pm1.341$ & $\pm1.468i$ & $\pm1.431i$ & saddle--centre--centre \\
  $E_4$ & $-0.673 \pm0.994i$ & $0.673 \pm0.994i$ & $\pm1.156i$ & sink--source--centre \\
  $E_5$ & $\pm2.330$ & $\pm5.752i$ & $\pm4.084i$ & saddle--centre--centre \\
  $E_6$ & $\pm4.568i$ & $\pm4.421i$ & $\pm1.982i$ & centre--centre--centre \\
  $E_7$ & $\pm5.280i$ & $\pm3.791i$ & $\pm1.447i$ & centre--centre--centre \\
  \hline
  \multicolumn{5}{c}{Large lobe} \\
  $L_1$ & $-1.059 \pm1.362i$ & $1.059 \pm1.362i$ & $\pm2.394i$ & sink--source--centre \\
  $L_2$ & $\pm2.875$ & $\pm3.103i$ & $\pm2.416i$ & saddle--centre--centre \\
  $L_3$ & $-1.280 \pm1.516i$ & $1.280 \pm1.516i$ & $\pm2.424i$ & sink--source--centre \\
  $L_4$ & $\pm2.424$ & $\pm2.825i$ & $\pm2.257i$ & saddle--centre--centre \\
  $L_5$ & $-4.493 \pm1.516i$ & $4.493 \pm1.680i$ & $\pm6.475i$ & sink--source--centre \\
  \hline
  \multicolumn{5}{c}{Small lobe} \\
  $S_1$ & $\pm2.103i$ & $\pm1.323i$ & $\pm1.013i$ & centre--centre--centre \\
  $S_2$ & $\pm1.169$ & $\pm2.099i$ & $\pm2.040i$ & saddle--centre--centre \\
  $S_3$ & $-0.842 \pm1.588i$ & $0.842 \pm1.588i$ & $\pm1.890i$ & sink--source--centre \\
  $S_4$ & $\pm1.549$ & $\pm2.253i$ & $\pm2.126i$ & saddle--centre--centre \\
  $S_5$ & $\pm9.830$ & $\pm7.623i$ & $\pm6.761i$ & saddle--centre--centre \\
  \hline
 \end{tabular}}
 \end{table}


\bsp	
\label{lastpage}
\end{document}